\documentclass[aps,prd,twocolumn,superscriptaddress,showpacs]{revtex4}
\usepackage{graphicx}
\usepackage{dcolumn}
\usepackage{bm}
\usepackage{natbib}

\newcommand{\aaps}{{Astron.~Astrophys.~Supp.}}
\newcommand{\araa}{{Annu.~Rev.~Astron.~Astrophys.}}
\newcommand{\aap}{{Astron.~Astrophys.}}
\newcommand{\apjl}{{Astrophys.~J.~Lett.}}
\newcommand{\apjs}{{Astrophys.~J.~Supp.}}
\newcommand{\aj}{{Astron.~J.}}
\newcommand{\mnras}{{Mon.~Not.~R.~Astron.~Soc.}}

\newcommand{\beq}{\begin{equation}}
\newcommand{\eeq}{\end{equation}}
\newcommand{\beqa}{\begin{eqnarray}}
\newcommand{\eeqa}{\end{eqnarray}}
\newcommand{\backten}{\!\!\!\!\!\!\!\!\!\!}
\newcommand{\backtwen}{\backten\backten}
\newcommand{\fwten}{$\;\;\;\;\;\;\;\;\;\;$}

\newcommand{\nhat}{\hat{\bf n}}

\newcommand{\WMAP}{{\slshape WMAP~}}
\newcommand{\WMAPc}{{\slshape WMAP}}

\newcommand{\Tr}{{\rm ~Tr~}}
\newcommand{\threej}[6]{{\left( \begin{array}{ccc} #1 & #2 & #3 \\ #4 & 
   #5 & #6 \end{array} \right)}}

\newcommand{\lmax}{{l_{\rm max}}}
\newcommand{\lmaxtwo}{{l_{\rm max}^2}}
\newcommand{\lmaxthree}{{l_{\rm max}^3}}

\newcommand{\isten}{$\cap$S10~}
\newcommand{\istenps}{$\cap$S10$\setminus$ps~}
\newcommand{\istenpst}{$\cap$S10$\setminus$ps$_2$~}

\begin{document}

\title{Cross-correlation of CMB with large-scale structure: weak 
gravitational lensing}

\author{Christopher M. Hirata}
\email{chirata@princeton.edu}
\affiliation{Department of Physics, Jadwin Hall, Princeton University, 
  Princeton, New Jersey 08544, USA}

\author{Nikhil Padmanabhan}
\affiliation{Department of Physics, Jadwin Hall, Princeton University, 
  Princeton, New Jersey 08544, USA}

\author{Uro\v s Seljak}
\affiliation{Department of Physics, Jadwin Hall, Princeton University, 
  Princeton, New Jersey 08544, USA}

\author{David Schlegel}
\affiliation{Princeton University Observatory, Princeton University,
  Princeton, New Jersey 08544, USA}

\author{Jonathan Brinkmann}
\affiliation{Apache Point Observatory, 2001 Apache Point Road, Sunspot,
  New Mexico 88349-0059, USA}

\date{August 30, 2004}

\begin{abstract}
We present the results of a search for gravitational lensing of the cosmic
microwave background (CMB) in cross-correlation with the projected density
of luminous red galaxies (LRGs).  The CMB lensing reconstruction is 
performed using the first year of {\slshape Wilkinson Microwave Anisotropy 
Probe} (\WMAPc) data, and the galaxy maps are obtained using the Sloan 
Digital Sky Survey (SDSS) imaging data. We find no detection of lensing; 
our constraint on the galaxy bias derived from the galaxy-convergence
cross-spectrum is $b_g=1.81\pm 1.92$ ($1\sigma$, statistical), as compared 
to the expected result of $b_g\sim 1.8$ for this sample.  We discuss 
possible instrument-related systematic errors and show that the Galactic 
foregrounds are not important.  We do not find any evidence for point 
source or thermal Sunyaev-Zel'dovich effect contamination.
\end{abstract}


\maketitle

\section{Introduction}
\label{sec:intro}

The {\slshape Wilkinson Microwave Anisotropy Probe} (\WMAPc)  
\footnote{URL: {\tt http://map.gsfc.nasa.gov/}} satellite has provided a
wealth of information about the universe through its high-resolution,
multi-frequency, all-sky maps of the cosmic microwave background (CMB)
\cite{2003ApJS..148....1B}.  While the \WMAP power spectrum
\cite{2003ApJS..148..135H} and temperature-polarization cross-spectrum
\cite{2003ApJS..148..161K} are useful for probing the high-redshift
universe (reionization and earlier epochs) \cite{2003ApJS..148..175S,
2003ApJS..148..195V, 2003ApJS..148..213P, 2003ApJS..148..233P}, the \WMAP
maps also provide an opportunity to study the low-redshift universe
through secondary CMB anisotropies. While the effect of secondary
anisotropies on the angular scales probed by \WMAP ($l\lesssim 700$) is
small compared to the primordial temperature fluctuations, the
signal-to-noise ratio can be boosted by cross-correlating with tracers of
the large scale structure (LSS) at low redshifts. Since the \WMAP data
release, several authors have used various tracers of LSS to measure the
integrated Sachs-Wolfe (ISW) effect, the thermal Sunyaev-Zel'dovich (tSZ)
effect, and microwave point sources \cite{2003MNRAS.346..940D,
2004Natur.427...45B, 2004ApJ...608...10N, 2004MNRAS.347L..67M,
2003ApJ...597L..89F, 2003astro.ph..7335S, 2004PhRvD..69h3524A}. The Sloan
Digital Sky Survey (SDSS) \footnote{URL: {\tt http://www.sdss.org/}} is an
excellent candidate for these cross-correlation studies due to the large
solid angle covered at moderate depth.

Another secondary anisotropy, which has not yet been investigated
observationally, is weak lensing of the CMB by intervening large-scale
structure.  Weak lensing has attracted much attention recently as a means
of directly measuring the matter power spectrum at low redshifts (e.g.
\cite{2003ARA&A..41..645R}).  The traditional approach is to use distant
galaxies as the ``sources'' that are lensed to measure e.g. the matter
power spectrum (e.g. \cite{2000A&A...358...30V, 2000MNRAS.318..625B,
2001ApJ...552L..85R, 2002ApJ...572...55H, 2002A&A...393..369V,
2003AJ....125.1014J, 2003MNRAS.341..100B, 2004astro.ph..4195M}) or the
galaxy-matter cross-correlation (e.g. \cite{1996ApJ...466..623B,
2000AJ....120.1198F, 2001astro.ph..8013M, 2002ApJ...577..604H,
2004AJ....127.2544S, 2004ApJ...606...67H}).  However weak lensing of CMB
offers an alternative method, free of intrinsic alignments, uncertainties
in the source redshift distribution, and selection biases (since the CMB
is a random field).  Potential applications of CMB lensing described in
the literature include precision measurement of cosmological parameters
\cite{2002ApJ...574..566H, 2002PhRvD..65b3003H, 2003NewAR..47..893K,
2004PhRvD..69d3004H} and separation of the lensing contribution to the CMB
$B$-mode polarization from primoridal vector \cite{2004PhRvD..70d3518L}
and tensor perturbations \cite{2002PhRvL..89a1303K, 2002PhRvL..89a1304K,
2004PhRvD..69d3005S}.  While these applications are in the future, the
\WMAP data for the first time allows a search for weak lensing of the CMB
in correlation with large-scale structure.  This paper presents the
results of such a search; our objective here is not precision cosmology,
but rather to detect and characterize any systematic effects that
contaminate the lensing signal at the level of the current data.  This
step is a prerequisite to future investigations that will demand tighter
control of systematics.

In this paper, we perform cross-correlation analysis between the CMB
weak lensing field derived from \WMAP and a photometrically selected
sample of luminous red galaxies (LRGs) in the SDSS at redshifts
$0.2<z<0.7$.  The photometric LRGs are well-suited for
cross-correlation studies because of their high intrinsic luminosity
(compared to normal galaxies), which allows them to be observed at
large distances; their high number density, which suppresses shot
noise in the maps; and their uniform colors which allow for accurate
photometric redshifts and hence determination of the redshift
distribution.  We use the measured cross-spectrum between the lensing
field and the projected galaxy density to estimate the LRG bias $b_g$.  
At the present stage, we are using the bias as a proxy for the
strength of the cross-correlation signal, just as has been done in
recent analyses of the ISW effect \cite{2004Natur.427...45B,
2004ApJ...608...10N, 2003ApJ...597L..89F, 2003astro.ph..7335S,
2004PhRvD..69h3524A}; we are not yet trying to use the bias in
cosmological parameter estimation, although this is a possible future
application of the methodology.  We do not have a detection of a
cross-correlation, and hence our measured bias $b_g=1.81\pm 1.92$ is
consistent with zero.

This paper is organized as follows.  The most important aspects (for this
analysis) of the \WMAP and SDSS data sets, and the construction of the LRG
catalog, are described in Sec.~\ref{sec:data}.  The theory of CMB lensing
and reconstruction methodology are explained in Sec.~\ref{sec:cmblens}.  
The cross-correlation methodology and simulations are covered in
Sec.~\ref{sec:crosscorrel}, and the results are presented in
Sec.~\ref{sec:results}.  We investigate possible systematic errors in
Sec.~\ref{sec:systematics}, and conclude in Sec.~\ref{sec:discussion}.
Appendix~\ref{app:nilsht} describes the spherical harmonic transform
algorithms and associated conventions used in this paper, and
Appendix~\ref{app:prec} describes the algorithm used for the ${\sf 
C}^{-1}$ operations that arise in our analysis.

\section{Data}
\label{sec:data}

\subsection{CMB temperature from \WMAPc}
\label{sec:wmap}

The \WMAP mission \cite{2003ApJ...583....1B} is designed to produce
all-sky maps of the CMB at multipoles up to $l\sim$several hundred.  
This analysis uses the first public release of \WMAP data, consisting
of one year of observations from the Sun-Earth L2 Lagrange point.  
\WMAP carries ten differencing assemblies (DAs), each of which
measures the difference in intensity of the CMB at two points on the
sky; a CMB map is buit up from these temperature differences as the
satellite rotates.  (\WMAP has polarization sensitivity but this is
not used in the present analysis.) The DAs are designated K1, Ka1, Q1,
Q2, V1, V2, W1, W2, W3, and W4; the letters indicate the frequency
band to which a particular DA is sensitive \cite{2003ApJS..148....1B,
2003ApJS..148...29J} (the K, Ka, Q, V, and W bands correspond to
central frequencies of 23, 33, 41, 61, and 94 GHz, respectively).  
The \WMAP team has pixelized the data from each DA in the HEALPix
\footnote{URL: {\tt http://www.eso.org/science/healpix/}} pixelization
system at resolution 9 \cite{2003ApJS..148....1B,
2003ApJS..148...63H}.  This system has 3,145,728 pixels, each of solid
angle 47.2 sq. arcmin. These maps are not beam-deconvolved; this,
combined with the \WMAP scan strategy, results in nearly uncorrelated
Gaussian uncertainties on the temperature in each pixel.

In this paper, we only use the three high-frequency microwave bands
(Q, V, and W) because the K and Ka bands are very heavily contaminated
by galactic foregrounds and have poor resolution.  (The foreground
emission is not a Gaussian field and cannot be reliably simulated, so
in cases where it dominates over CMB anisotropy and instrument noise,
we cannot compute reliable error bars on the cross-correlation.)  For
the galaxy-lensing correlation, we have used the sky maps produced by
the eight high-frequency DAs. The variances of the temperature
measurements are obtained from the effective number of observations
$N_{obs}$.

Note that the \WMAP ``internal linear combination'' (ILC) map
\cite{2003ApJS..148....1B} cannot be used for lensing studies because
of its degraded resolution (1 degree full width half maximum, FWHM),
which eliminates the multipoles $l\sim 350$ of greatest importance for
the lensing analysis.  An ILC-based lensing analysis would also suffer
from practical issues, namely the loss of frequency-dependent
information (useful as a test of foregrounds), the inability to
separate cross-correlations between different DAs from
auto-correlations (useful to avoid the need for noise bias
subtraction), and the complicated inter-pixel noise correlations (due
to the smoothing used to create the map and the varying weights of the
different frequencies).  The foreground-cleaned map of
Ref.~\cite{2003PhRvD..68l3523T} recovers the full \WMAP resolution,
but the practical difficulties (for the purpose of lensing
reconstruction) associated with ILC still apply.  We have not used
either of these maps in this paper.

\subsection{LRG density from SDSS}

The Sloan Digital Sky Survey (SDSS) \cite{2000AJ....120.1579Y} is an
ongoing survey to image approximately $\pi$ steradians of the sky, and
follow up approximately one million of the detected objects
spectroscopically \cite{2002AJ....124.1810S, 2002AJ....123.2945R}. The
imaging is carried out by drift-scanning the sky in photometric conditions
\cite{2001AJ....122.2129H}, in five bands ($ugriz$)
\cite{1996AJ....111.1748F, 2002AJ....123.2121S} using a specially designed
wide-field camera \cite{1998AJ....116.3040G}. These imaging data are the
source of the LSS sample that we use in this paper. In addition, objects
are targeted for spectroscopy using these data \cite{2003AJ....125.2276B}
and are observed with a 640-fiber spectrograph on the same telescope. All
of these data are processed by completely automated pipelines that detect
and measure photometric properties of objects, and astrometrically
calibrate the data \cite{2001adass..10..269L, 2003AJ....125.1559P}. The
SDSS is well underway, and has had three major data releases
\cite{2002AJ....123..485S, 2003AJ....126.2081A, 2004AJ....128..502A,
dfink}; this paper uses all data observed through Fall 2003 (296,872
HEALPix resolution 9 pixels, or 3,893 square degrees).

The SDSS detects many extragalactic objects that could, in principle,
be used for cross-correlation with secondary anisotropies
\cite{2000ApJ...540..605P}.  The usefulness of LRGs as a cosmological
probe has been appreciated by a number of authors
\cite{2000AJ....120.2148G, 2001AJ....122.2267E}. These are typically
the most luminous galaxies in the universe, and therefore probe
cosmologically interesting volumes.  In addition, these galaxies are
generically old stellar systems and have extremely uniform spectral
energy distributions (SEDs), characterized only by a strong
discontinuity at 4000~\AA. The combination of these two
characteristics make them an ideal candidate for photometric redshift
algorithms, with redshift accuracies of $\sigma_z \sim 0.03$
\cite{2004astro.ph..7594P}. We briefly outline the construction of the
photometric LRG sample used in this paper below, and defer a detailed
discussion of the selection criteria and properties of the sample to a
later paper \citep{padauto}.

Our selection criteria are derived from those described in
Ref.~\cite{2001AJ....122.2267E}. However, since we are working with a 
photometric
sample, we are able to relax the apparent luminosity constraints
imposed there to ensure good throughput on the SDSS spectrographs. We 
select
LRGs by choosing galaxies that both have colors consistent with
an old stellar population, as well as absolute luminosities greater
than a chosen threshold. The first criterion is simple to implement since
the uniform SEDs of LRGs imply that they lie on an extremely tight locus
in the space of galaxy colors; we simply select all galaxies that lie
close to that locus. More specifically, we can define three (not 
independent) colors that describe this locus,
\beqa
c_{\perp} &\equiv & (r-i) - 0.25(g-r) - 0.18 \,\,\, , \nonumber \\
d_{\perp} &\equiv & (r-i) - 0.125(g-r) \,\,\,, \nonumber \\
c_{||} &\equiv & 0.7 (g-r) + 1.2(r-i-0.18) \,\,\,,
\label{eq:perpdef}
\eeqa
where $g$, $r$, and $i$ are the SDSS model magnitudes 
\cite{2002AJ....123..485S}
in the $g, r$ and $i$ bands (centered at 469, 617, and 748~nm 
respectively).  We now make the following color selections,
\beqa
{\rm Cut\,\,I :} & \mid c_{\perp} \mid < 0.2; \nonumber \\
{\rm Cut\,\,II :} & d_{\perp} > 0.55, \,\,\, g-r > 1.4.
\label{eq:colourcuts}
\eeqa
Making two cuts (Cut I and Cut II) is convenient since the LRG
color locus changes direction sharply as the 4000 \AA~break
redshifts from the $g$ to the $r$ band; this division divides the
sample into a low redshift (Cut I, $z < 0.4$) and high
redshift (Cut II, $z > 0.4$) sample.

In order to implement the absolute magnitude cut, we follow
\cite{2001AJ....122.2267E} and impose a cut in the
galaxy color-magnitude space. The specific cuts we use are
\beqa
{\rm Cut\,\,I :} && r_{Petro} < 13.6 + \frac{c_{||}}{0.3}, \,\,\,
r_{Petro} < 19.7,
\nonumber \\
{\rm Cut\,\,II :} && i < 18.3 + 2d_{\perp}, \,\,\, i < 20,
\eeqa
where $r_{Petro}$ is the SDSS $r$ band Petrosian magnitude 
\cite{2002AJ....123..485S}.  Finally, we reject all objects that resemble 
the point-spread function of the telescope, or if they have colors 
inconsistent with normal galaxies; these cuts attempt to remove 
interloping stars.

\begin{figure}
\includegraphics[width=3in]{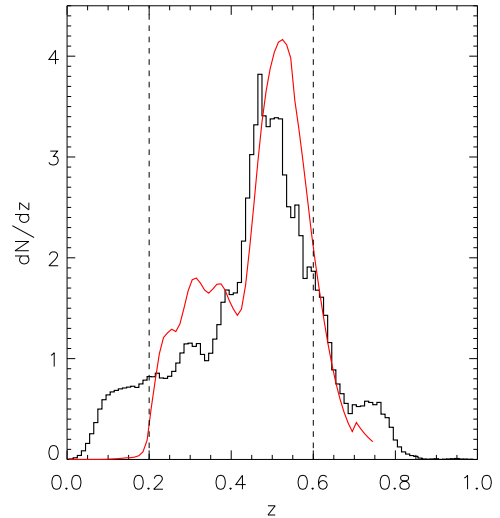}
\caption{\label{fig:lrgdndz}The LRG redshift distribution.  The black 
histogram shows the photo-$z$ distribution, the red curve is the true 
redshift distribution estimated by regularized deconvolution of the 
photo-$z$ errors.}
\end{figure}

Applying these selection criteria to the $\sim $ 5500 degress of
photometric SDSS imaging in the Galactic North yields a catalog of
approximately 900,000 galaxies. Applying the single template fitting
photometric redshift algorithm of \cite{2004astro.ph..7594P}, we restrict
this catalog to galaxies with $0.2 < z_{photo} < 0.6$, leaving us with
$\sim$ 650,000 galaxies. We use the regularized inversion method of
Ref.~\cite{2004astro.ph..7594P} as well as the photometric redshift error
distribution presented there, to estimate the true redshift distribution
of the sample. The results, comparing the photometric and true redshift
distributions are shown in Fig.~\ref{fig:lrgdndz}. Finally, this catalog
is pixelized as a number overdensity, $g=\delta n/\bar n$, onto a HEALPix
pixelization of the sphere, with 3,145,728 pixels. We also mask regions
around stars from the Tycho astrometric catalog
\cite{2000A&A...355L..27H}, as the photometric catalogs are incomplete
near bright stars. The final catalog covers a solid angle of 3,893 square
degrees (296,872 HEALPix resolution 9 pixels)  and contains 503,944
galaxies at a mean density of $1.70$ galaxies per pixel.

\section{Lensing of CMB}
\label{sec:cmblens}

\subsection{Definitions}
\label{sec:not}

Gravitational lensing re-maps the primordial CMB anisotropy $\tilde T$
into a lensed temperature $T$ according to
\beq
T(\nhat) = \tilde T(\nhat + {\bf d}(\nhat)),
\label{eq:t-tilde}
\eeq
where the 2-vector ${\bf d}$ is the deflection angle of null geodesics.  
To first order in the metric perturbations, ${\bf d}$ can be expressed as
the gradient of a scalar lensing potential, ${\bf d}=\nabla\Phi$, where
$\nabla$ is the derivative on the unit (celestial) sphere.  We may also
define the convergence $\kappa=-{1\over 2}\nabla\cdot{\bf d}$.  Assuming
the primordial CMB is statistically isotropic with some power spectrum
$C_l$, it can be shown \cite{2001PhRvD..64h3005H,2003PhRvD..67h3002O} that
the multipole moments of the lensed temperature field have covariance
\beqa
\langle T_{l_1m_1}^\ast T_{l_2m_2} \rangle &=& 
C_{l_1}\delta_{l_1l_2}\delta_{m_1m_2} +
\sum_{LM} (-1)^{m_2} {\cal J}_{Ll_1l_2} 
\nonumber \\ && \times
\threej{l_1}{l_2}{L}{-m_1}{m_2}{-M} \kappa_{LM},
\label{eq:tcov}
\eeqa
where we have introduced the Wigner $3j$ symbol, and the coupling 
coefficient is
\beqa
{\cal J}_{Ll_1l_2} &=& {2\over L(L+1)} \sqrt{(2L+1)(2l_1+1)(2l_2+1)\over 16\pi}
\nonumber \\ && \times \bigl\{ [L(L+1) + l_1(l_1+1) - l_2(l_2+1)]C_{l_1} 
\nonumber \\ && + [L(L+1) - l_1(l_1+1) + l_2(l_2+1)]C_{l_2} \bigr\}
\nonumber \\ && \times \threej{l_1}{l_2}{L}{0}{0}{0}.
\label{eq:jlll}
\eeqa

The \WMAP satellite does not directly measure $T$, but rather a
beam-convolved temperature:
\beq
\dot T^\alpha(\nhat) = \int B^\alpha(\nhat, \nhat') T(\nhat') d^2\nhat' + 
{\rm noise}.
\eeq
(Here $\alpha$=K1..W4 are the differencing assemblies.) In most of this
analysis we have approximated the beam by the \WMAP circularized beam
transfer function \cite{2003ApJS..148...39P}.  The temperature multipole
moments recovered assuming a circular beam are
\beq
\hat T^\alpha_{lm} = \frac{ \dot T^\alpha_{lm} }{ B_l^\alpha },
\label{eq:that}
\eeq
where $B_l^\alpha$ are the beam transfer functions.  If the beam is truly
circular, Eq.~(\ref{eq:that}) returns an unbiased estimator of the
beam-deconvolved CMB temperature; in Sec.~\ref{sec:beam}, we will consider
the effect of the \WMAP beam ellipticity on lensing estimation.  Note
further that $\hat T^\alpha_{lm}$ is only well-determined up to some
maximum multipole $l$ because the $B_l^\alpha$ drop to zero at high $l$.

\subsection{Theoretical predictions for lensing}
\label{sec:theory}

In this paper, we aim to measure the galaxy-convergence cross-correlation,
$C^{g\kappa}_l$, where $g=\delta n/\bar n$ is the projected fractional
overdensity of galaxies; this section briefly presents the theoretical
prediction from the $\Lambda$CDM cosmology.  In a spatially flat
Friedmann-Robertson-Walker universe described by general relativity, the
convergence is given in terms of the fractional density perturbation
$\delta$ by:
\beq
\kappa(\nhat) = 4\pi G_N\bar\rho_0 \int 
\frac{\chi(\chi_{CMB}-\chi)}{\chi_{CMB}} (1+z) \delta(\chi,\nhat) d\chi,
\eeq
where $\chi$ is the comoving radial distance, $z$ is the redshift observed
at radial distance $\chi$, $\bar\rho_0$ is the present-day mean density of
the universe, and $\chi_{CMB}$ is the comoving distance to the CMB.  The
galaxy overdensity does not come from a ``clean'' theoretical prediction,
but on large scales it can be approximated by
\beq
g(\nhat) = \frac{ \int b_g {\cal N}(\chi) \delta(\chi,\nhat) d\chi }{ 
\int {\cal N}(\chi) d\chi },
\label{eq:gnhat}
\eeq
where ${\cal N}(\chi)$ is the distribution in comoving distance and $b_g$
is the galaxy bias.  (The SDSS LRG sample is at low redshift $z\le 0.7$ 
and does not have a steep luminosity function at the faint end of our 
sample, so we neglect the magnification bias.)  For $l\gg 1$ and smooth 
power spectra for matter and galaxies, this may be approximated by the 
Limber integral:
\beqa
C^{g\kappa}_l &=& 4\pi G_N\bar\rho_0
\nonumber \\ && \times
\frac{
\int b_g{\cal N}(\chi) ({1\over\chi}-{1\over\chi_{CMB}}) (1+z) 
P_\delta(\frac{l}{\chi}) d\chi}{ \int {\cal N}(\chi) d\chi }, \;\;\;
\label{eq:limber}
\eeqa
where the matter power spectrum $P_\delta$ is evaluated at comoving
wavenumber $k=l/\chi$ and at the redshift corresponding to conformal
time $\eta_0-\chi$.  It is obtained using the transfer functions from
{\sc CMBFast} \cite{1996ApJ...469..437S} and the best-fit 6-parameter
flat $\Lambda$CDM cosmological model from \WMAP and SDSS data from
Ref.~\cite{2004PhRvD..69j3501T} ($\Omega_bh^2=0.0232$;  
$\Omega_mh^2=0.1454$; $h=0.695$; $\tau=0.124$; $\sigma_8=0.917$;  
$n_s=0.977$).  We have found that varying each of these parameters
over their $1\sigma$ uncertainty ranges gives a $\pm 18\%$ effect for
$\sigma_8$ (for which $C^{g\kappa}_l$ scales as $\propto\sigma_8^2$)  
and $\pm<14\%$ effect for the other parameters.  Since the overall
significance of the detection is only 0.9$\sigma$, this dependence of
the template $C^{g\kappa}_l$ on cosmological parameters will be
neglected here.

The lensing signal is on large scales and so we have not used a
nonlinear mapping.  The peak of the LRG redshift distribution is at
$z\sim 0.5$, corresponding to a comoving angular diameter distance
$\sim 1.3h^{-1}$~Gpc, in which case the smallest angular scales we use
($l=300$) correspond to $k=0.23h$~Mpc$^{-1}$ and $\Delta^2(k)=0.7$.  
The nonlinear evolution at this scale according to the Peacock-Dodds
formula \cite{1996MNRAS.280L..19P} is a $10\%$ correction to the
matter power spectrum and is thus much smaller than the error bars
presented in this paper (although it is not obvious what this implies
about the galaxy-matter cross-spectrum, which is the quantity that
should appear in Eq.~\ref{eq:limber}).  Future applications of CMB
lensing in precision cosmology will of course require accurate
treatment of the nonlinear evolution.

In this paper, we will assume that galaxy bias $b_g$ is constant so that
it may be pulled out of the integrals in Eqs.~(\ref{eq:gnhat}) and
(\ref{eq:limber}).  If the bias varies with redshift (as suggested by
e.g. \cite{1996ApJ...461L..65F}), then the best-fit value of $b_g$ will
be some weighted average of $b_g$ over the redshift distribution; this
need not be the same weighted average that one observes from the
auto-power spectrum, since the latter is weighted differently.  
Computation of the auto-power in photometric redshift slices
\cite{padauto} suggests that over the redshift range $0.2\le z\le 0.6$
the bias varies from $1.7$--$1.9$; this variation can safely be
neglected given our current statistical errors.  Note however that a
detection of $b_g\neq 0$ via a fit to Eq.~(\ref{eq:limber}) assuming
constant $b_g$ would rule out $C_l^{g\kappa}=0$ and hence would be
robust evidence for a galaxy-convergence correlation, regardless of the
redshift dependence of the bias.

\subsection{Lensing reconstruction}
\label{sec:lensing}

We construct lensing deflection maps using quadratic reconstruction
methods \cite{1997A&A...324...15B, 1999PhRvL..82.2636S,
2000PhRvD..620a9fdG, 2000PhRvD..62f3510Z, 2001ApJ...557L..79H,
2002PhRvD..66f3008O, 2002ApJ...574..566H, 2003PhRvD..67h3002O,
2004NewA....9..173C}, which have been shown to be near-optimal for lensing
studies of the CMB temperature on large ($l<3500$) scales
\cite{2003PhRvD..67d3001H}.  Non-quadratic methods may be superior if CMB
polarization is used \cite{2003PhRvD..68h3002H}, or on very small scales
\cite{2000ApJ...538...57S, 2004astro.ph..2004V, 2004PhRvD..70b3009D};  
these cases are not of interest for \WMAPc, since the sensitivity is
insufficient to map the CMB polarization and arcminute scales are
unresolved.  Quadratic estimation takes advantage of the cross-coupling of
different multipoles induced by gravitational lensing, namely the
$O(\kappa_{LM})$ term in Eq.~(\ref{eq:tcov}). The maximum signal-to-noise
statistic for CMB weak lensing is the divergence of the
temperature-weighted gradient. We construct, for each pair of differencing
assemblies $\alpha$ and $\beta$, the temperature-weighted gradient vector
field $\tilde{\bf G}^{\alpha\beta}$:
\beqa
\tilde{\bf G}^{\alpha\beta}(\nhat) &=& {1\over 2} \biggl( [W\hat 
T^\alpha](\nhat)\nabla[CW\hat T^\beta](\nhat)
\nonumber \\ && + [W\hat T^\beta](\nhat)\nabla[CW\hat T^\alpha](\nhat) \biggr),
\label{eq:gdef}
\eeqa
where $W\hat T$ and $CW\hat T$ are defined by the convolution relations
$[W\hat T^\alpha]_{lm}=W_l\hat T^\alpha_{lm}$ and $[CW\hat
T^\alpha]_{lm}=C_lW_l\hat T^\alpha_{lm}$.  Note that Eq.~(\ref{eq:gdef})
is exactly the same as the $\tilde{\bf G}$ statistic of
Ref.~\cite{2001ApJ...557L..79H} except that we have multiple differencing
assemblies, and we have left open the choice of the weight $W_l$.  While
$W_l=(C_l+C_l^{noise})^{-1}$ is statistically optimal, there are also
practical considerations that affect this choice. Specifically, it is
desirable that the $W\hat T$ and $CW\hat T$ convolutions are almost-local
functions of the CMB temperature (to minimize leakage from the Galactic
Plane), and that the same $W_l$ be used for all differencing assemblies
(so that any frequency dependence of our results can be attributed to
foregrounds or noise, rather than merely a change in which primary CMB
modes we are studying).  We choose the following weight function:
\beq
W_l = \left[ C_l + (0.03461\mu{\rm K}^2)e^{l(l+1)/300^2} \right]^{-1},
\label{eq:wl}
\eeq
which clearly has the optimal $C_l^{-1}$ dependence in the high
signal-to-noise regime.  Note that in the range of $l$ we use ($l\le
800$), the $W_l$ drop to zero with increasing $l$ faster than the Q-band
beam transfer functions.  Hence the computation of $W\hat
T^{Q1,Q2}$ are stable even though $\hat T^{Q1,Q2}$ are beam-deconvolved.  
Due to their narrower beams, this stability also applies to the V and
W-band DAs.  The K1 and Ka1 DAs have wider beams and hence lensing
reconstruction using the weight Eq.~(\ref{eq:wl}) is unstable for these
DAs.  We set $W_0=W_1=0$ to reject the monopole (not observed by \WMAPc)  
and dipole signals.  The power spectrum $C_l$ used for the lensing
reconstruction is \WMAP best-fit $\Lambda$CDM model with scalar 
spectral running $\alpha_s$
\cite{2003ApJS..148..175S} to the CMB data (\WMAPc +ACBAR
\cite{2002AAS...20114004K} +CBI \cite{2002AAS...200.0606P}).  Errors in
the $C_l$ used in the lensing reconstruction cannot produce a spurious
galaxy-temperature correlation because they result only in a calibration
error in the lensing estimator.  Furthermore, \WMAP has determined the 
$C_l$ to
within several percent (except at the low multipoles, which give a 
subdominant contribution to both $W\hat T$ and $\nabla CW\hat T$), whereas 
the lensing cross-correlation signal is only present at the $1\sigma$ 
level, this error is not important for the present analysis.

In the reconstruction method of Ref.~\cite{2001ApJ...557L..79H}, a
filtered divergence of $\tilde{\bf G}$ is taken to extract the lensing
field.  We avoid this step because it is highly nonlocal, and hence can
smear Galactic plane contamination into the regions of sky used for
lensing analysis.  In principle, we would prefer to directly
cross-correlate $\tilde{\bf G}$ with the LRG map, but this too is
difficult because the noise power spectrum of $\tilde{\bf G}$ is extremely
blue.  We compromise by computing ${\bf v}$, a Gaussian-filtered version
of $\tilde{\bf G}$:
\beq
v_{lm}^{(\parallel,\perp)} = e^{-l(l+1)\sigma_0^2/2}\tilde 
G_{lm}^{(\parallel,\perp)}.
\label{eq:vdef}
\eeq
Here $\parallel$ and $\perp$ represent the longitudinal (vector) and
transverse (axial) multipoles, which are the vector analogues of the
tensor $E$ and $B$ multipoles.  The Gaussian filter eliminates the troublesome
high-$l$ power present in $\tilde{\bf G}(\nhat)$ and makes ${\bf v}$
suitable for cross-correlation studies.  We have chosen a width 
$\sigma_0=0.01$ radians (34 arcmin).

\begin{figure}
\includegraphics[angle=-90,width=3in]{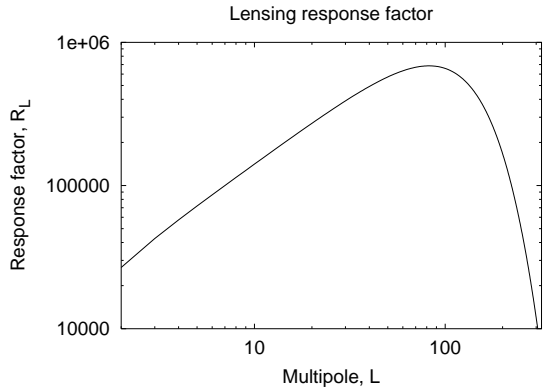}
\caption{\label{fig:rl}
The response factor $R_l$ of Eq.~(\ref{eq:calib}) satisfying $\langle 
v^{\alpha\beta(\parallel)}_{lm}\rangle = R_l\kappa_{lm}$.
}
\end{figure}

The vector field ${\bf v}$ can be written in terms of the temperatures 
$\hat T$ directly in harmonic space.  The longitudinal components 
are given by
\beqa
v_{lm}^{\alpha\beta (\parallel)} &=& (-1)^m \sum_{l'l''} {\cal 
K}_{ll'l''} \sum_{m'm''} \threej{l}{l'}{l''}{-m}{m'}{m''} 
\nonumber \\ && \times
{\hat T^\alpha_{l'm'} \hat T^\beta_{l''m''} + \hat T^\beta_{l'm'} \hat 
T^\alpha_{l''m''} \over 2},
\label{eq:harmonicv}
\eeqa
where we have defined:
\beqa
{\cal K}_{ll'l''} &=& \threej{l}{l'}{l''}{0}{0}{0} 
\sqrt{(2l+1)(2l'+1)(2l''+1)\over 16\pi l(l+1)}
\nonumber \\ && \times [ l(l+1) - l'(l'+1) + l''(l''+1) ] C_{l''} W_{l'} 
W_{l''}
\nonumber \\ && \times e^{-l(l+1)\sigma_0^2/2}.
\label{eq:klll}
\eeqa
We will not need the formula for the transverse components.  While we
perform the reconstruction (Eqs.~\ref{eq:gdef} and \ref{eq:vdef}) in
real space, the harmonic-space relation (Eq.~\ref{eq:harmonicv}) is useful
for computing the response of the estimator, and for estimating foreground
contamination and beam effects.  In particular, from the orthonormality
relations for Wigner $3j$-symbols, we have
\beq
\langle v^{\alpha\beta(\parallel)}_{lm} \rangle = \sum_{l'l''} {{\cal 
J}_{ll'l''}{\cal K}_{ll'l''} \over 2l+1} \kappa_{lm}
\equiv R_l \kappa_{lm},
\label{eq:calib}
\eeq
which defines the calibration of ${\bf v}$ as an estimator of the lensing
field.  The response factor $R_l$ is shown in Fig.~\ref{fig:rl}; we have 
verified this response factor in simulations (Sec.~\ref{sec:sims}).

There are 36 pairs of differencing assemblies that could be used to
produce estimated lensing maps ${\bf v}^{\alpha\beta}_{lm}$: the 8
``autocorrelations'' ($\alpha=\beta$) and 28 ``cross-correlations''
($\alpha\neq\beta$).  Note that instrument noise that is non-uniform
across the sky can produce a bias in the ``autocorrelation''-derived
lensing map ${\bf v}^{\alpha\alpha}$.  In principle the noise bias could
be estimated and subtracted, just as can be done for the power spectrum.
However, this is dangerous if the noise properties are not very well
modeled. Since the \WMAP noise is in fact strongly variable across the sky
we only use the ``cross-correlation'' $\alpha\neq\beta$ maps.

One problem that we find with this method is that the ${\bf v}$ field
contains ``ghosts'' caused by the Galactic plane (where small-scale
temperature fluctuations of several millikelvin or more can occur due to
Galactic emission).  We solve this problem by setting $T=0$ within the
\WMAP Kp4 \cite{2003ApJS..148...97B} Galactic plane mask.  We have
verified that using the Kp2 mask instead produces only small changes to
the results.

\begin{figure}
\includegraphics[angle=-90,width=3in]{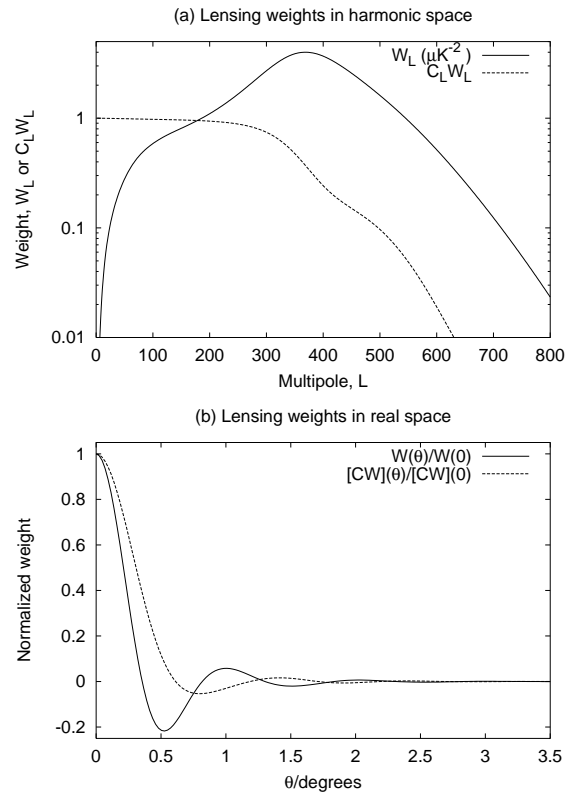}
\caption{\label{fig:cwplot}(a) The weight functions $W_l$ and $C_lW_l$.  
(b) The same weight functions in real-space 
(Eq.~\ref{eq:real}); $W\hat T$ and $CW\hat T$ are obtained by convolving 
the (beam-deconvolved) temperature $\hat T$ with these 
kernels.}
\end{figure}

The weight functions $W_l$ and $C_lW_l$ are shown in 
Fig.~\ref{fig:cwplot}.  We also show the real-space weights, given by
\beq
W(\theta) = \sum_l {2l+1\over 4\pi} W_l P_l(\cos\theta),
\label{eq:real}
\eeq
and similarly for $[CW](\theta)$.

\subsection{Frequency-averaged lensing maps}
\label{sec:falm}

\begin{figure}
\includegraphics[width=3.2in]{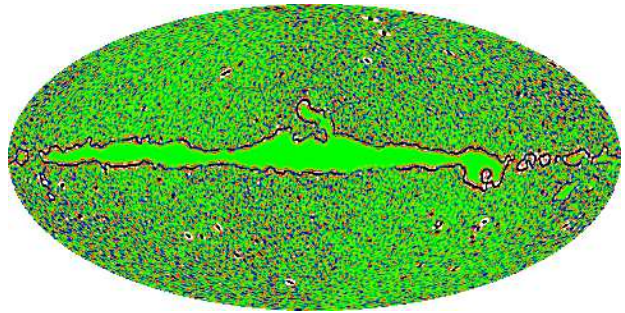}
\caption{\label{fig:divvtt}The divergence of the lensing vector field map,
$\nabla\cdot{\bf v}^{(TT)}$, smoothed with a 30 arcmin FWHM Gaussian, and
displayed in Galactic Molleweide projection.  Note the prominent artifacts 
surrounding the Galactic plane cut and the point sources (which are 
removed by the Kp05\istenpst cut).}
\end{figure}

The methodology outlined in Sec.~\ref{sec:lensing} allows us to construct
28 lensing maps ${\bf v}^{\alpha\beta}$ corresponding to the 28 pairs of 
differencing assemblies.  For this analysis, we
need to produce an ``averaged'' lensing map ${\bf v}^{(TT)}$ based on a
minimum-variance linear combination of the 28 DA-pair maps. The averaged
lensing map is determined by the weights $a^{(TT)}_{\alpha\beta}$:
\beq
{\bf v}^{(TT)} = \sum_{\alpha\beta} a^{(TT)}_{\alpha\beta} {\bf 
v}^{\alpha\beta}.
\eeq
We select these weights to minimize the amount of power in ${\bf
v}^{(TT)}$, subject to the restriction $\sum a^{(TT)}_{\alpha\beta}=1$;
this is done by minimizing the total vector power in ${\bf v}$ between
multipoles 50 and 125:
${\cal P} = \sum_{l=50}^{125} \sum_{m=-l}^l |v^{(TT)\,\parallel}_{lm}|^2$,
which is a quadratic function of the weights $a^{(TT)}_{\alpha\beta}$.  
The optimal weights $a^{(TT)}_{\alpha\beta}$ are complicated to establish
analytically since the maps ${\bf v}^{\alpha\beta}$ are highly correlated.  
We have therefore minimized ${\cal P}$ using a {\em simulated} lensing map
(see Sec.~\ref{sec:sims}).  Using a simulated map rather than the real 
data avoids the undesirable possibility of the weights being statistically
correlated with the data.  We also fix $a^{(TT)}_{Q1,Q2}=0$ because
the ${\bf v}^{Q1,Q2}$ map would be the most heavily contaminated by point 
sources.  The weights so obtained are shown in Table~\ref{tab:a}.  A map 
of $\nabla\cdot{\bf v}^{(TT)}$, smoothed to 30 arcmin resolution (Gaussian 
FWHM) is shown in Fig.~\ref{fig:divvtt}.

Also, to study foreground effects on lensing estimation, we would like to
construct averaged lensing maps ${\bf v}^{(QQ)}$, ${\bf v}^{(QV)}$, etc.
where we only average over differencing assemblies at the same
frequency, thereby preserving frequency-dependent information.  There are
six of these maps (QQ, QV, QW, VV, VW, and WW); the last
column of Table~\ref{tab:a} shows the weights used to construct them.  

\begin{table}
\caption{\label{tab:a}The weights $a_{\alpha\beta}$ used for the 
overall-averaged lensing map $a^{(TT)}_{\alpha\beta}$, and the 
six ``individual frequency'' maps 
$a^{(QQ)}_{\alpha\beta},...,a^{(WW)}_{\alpha\beta}$.}
\begin{tabular}{ccdcd}
\hline\hline
DA pair ($\alpha\beta$) & & a^{(TT)}_{\alpha\beta}\!\!\!\!\!\!\!\! & & 
a^{(\nu_1\nu_2)}_{\alpha\beta}\!\!\!\!\!\!\!\! \\
\hline
Q1,Q2 & & 0.000000 & & 1.000000 \\
\hline
Q1,V1 & & 0.063759 & & 0.220605 \\
Q1,V2 & & 0.075370 & & 0.267859 \\
Q2,V1 & & 0.061538 & & 0.227341 \\
Q2,V2 & & 0.098884 & & 0.284196 \\
\hline
Q1,W1 & & 0.031583 & & 0.167730 \\
Q1,W2 & & 0.026896 & & 0.113943 \\
Q1,W3 & & 0.018391 & & 0.094086 \\
Q1,W4 & & 0.028571 & & 0.130429 \\
Q2,W1 & & 0.023816 & & 0.158925 \\
Q2,W2 & & 0.017153 & & 0.107379 \\
Q2,W3 & & 0.014494 & & 0.097062 \\
Q2,W4 & & 0.027164 & & 0.130444 \\
\hline
V1,V2 & & 0.106330 & & 1.000000 \\
\hline
V1,W1 & & 0.053278 & & 0.150880 \\
V1,W2 & & 0.029381 & & 0.094494 \\
V1,W3 & & 0.029664 & & 0.087425 \\
V1,W4 & & 0.035463 & & 0.107193 \\
V2,W1 & & 0.048598 & & 0.168035 \\
V2,W2 & & 0.049979 & & 0.137365 \\
V2,W3 & & 0.034621 & & 0.115438 \\
V2,W4 & & 0.046029 & & 0.139169 \\
\hline
W1,W2 & & 0.016453 & & 0.199003 \\
W1,W3 & & 0.014448 & & 0.170408 \\
W1,W4 & & 0.019908 & & 0.232958 \\
W2,W3 & & 0.014514 & & 0.129381 \\
W2,W4 & & 0.003365 & & 0.133767 \\
W3,W4 & & 0.010347 & & 0.134484 \\
\hline\hline
\end{tabular}
\end{table}

\begin{figure*}
\includegraphics[angle=-90,width=5.5in]{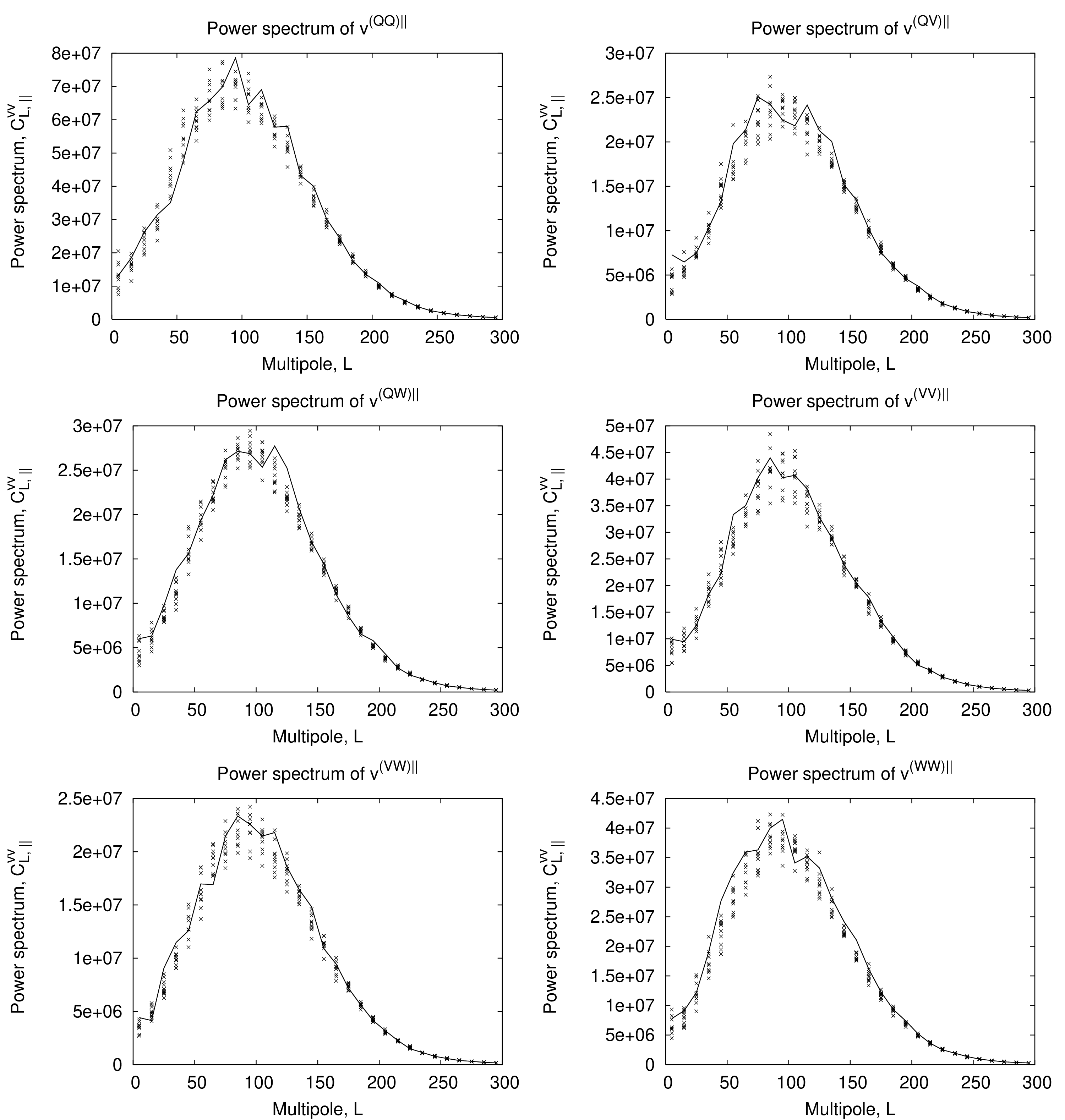}
\caption{\label{fig:vpower}The longitudinal mode power spectrum
$C_{l,\parallel}^{vv}$ within the Kp05\istenpst cut (solid line), for each
of the six frequency pairs.  The points show 10 simulations
(containing no lensing or foregrounds).  Since the purpose of this plot is
to compare the simulations to the actual data, the sky cut has not been
deconvolved, rather we have plotted the average of $|\int {\bf v}\cdot{\bf
Y}^{(\parallel)\ast}_{lm} d^2\nhat/f_{sky}|^2$ within bands of width
$\Delta l=10$.}
\end{figure*}

The power spectrum of the longitudinal mode of ${\bf v}$ obtained on the
cut sky (Kp05\istenpst cut, which excludes point sources; see
Sec.~\ref{sec:skycut}) is shown in Fig.~\ref{fig:vpower}.

\section{Cross-correlation computation}
\label{sec:crosscorrel}

\subsection{Sky cuts}
\label{sec:skycut}

In some regions of the sky, particularly the Galactic plane, microwave
emission from within the Milky Way and from nearby galaxies dominates over
the cosmological signal.  For their CMB analysis, the \WMAP team removed
this signal by (i) masking out a region based on a smoothed contour of the
K-band temperature, which they denote ``Kp2'' \cite{2003ApJS..148...97B},
and (ii) projecting out of their map microwave emission templates for
synchrotron, free-free, and dust emission based on other observations
\cite{1981A&A...100..209H, 1982A&AS...47....1H, 2003ApJS..146..407F,
1998ApJ...500..525S, 1999ApJ...524..867F}.  Template projection is
dangerous for cross-correlation studies involving galaxies because the
dust template of Ref.~\cite{1998ApJ...500..525S} is used to
extinction-correct the LRG magnitudes, thus template errors could
introduce spurious correlations between the CMB and galaxy maps.  Since
visual inspection of the uncleaned \WMAP maps reveals Galactic
contamination outside the Kp2-rejected region at all five frequencies, we
have used the more conservative Kp0 mask in Sec.~\ref{sec:extra} for our
galaxy-temperature correlations.  Because the SDSS covers only a small
portion of the sky, we speed up the cross-correlation computation by using
only \WMAP data in the vicinity of the SDSS survey region.  We define the
``S10'' region to consist of those pixels within 10 degrees of the SDSS
survey area.  The Kp0\isten cut accepts 774,534 HEALPix pixels (10,157 sq.
deg.).

When analyzing primary CMB anisotropies, it is customary to mask detected
point sources in order to eliminate this spurious contribution to the
temperature.  For secondary anisotropy studies, the analysis should be
done both with and without the point sources because the point sources may
correlate with large scale structure, hence naively masking them could
lead to misleading results. Therefore for the galaxy-temperature
correlation used in Sec.~\ref{sec:extra} we have constructed the
Kp0\istenps mask by rejecting all pixels within the \WMAP point source
mask (with a 0.6 degree exclusion radius around each source).  The
Kp0\istenps cut accepts 756,078 HEALPix pixels (9915 sq. deg.).

For the lensing analysis, we must use a more conservative mask than Kp0
because the lensing estimator ${\bf v}$ is a nonlocal function of the CMB
temperature, hence ${\bf v}(\nhat)$ responds to foreground emission
several degrees away from $\nhat$.  We have therefore constructed a
``Kp05'' mask consisting of all pixels within Kp0 that are at least 5
degrees away from the Kp0 boundary; the Kp05\isten mask used for the
lensing analysis accepts 753,242 HEALPix pixels (9,878 sq. deg.).  We have
also constructed a point source-removed version, Kp05\istenpst, in which
all pixels within 2 degrees of the point sources are rejected.  This mask
accepts 598,795 HEALPix pixels (7,853 sq. deg.).

\subsection{Galaxy-convergence correlation}
\label{sec:gk}

Having constructed the vector field ${\bf v}$, we proceed to compute its
cross-correlation $C^{gv}_l$ with the LRG map.  We construct the data 
vector
\beq
{\bf x}^T = ( {\bf x}^T_{LRG}, {\bf x}^T_{lens})
\eeq
of length $N_{pix}^{(LRG)}+2N_{pix}^{(CMB)}$, where ${\bf x}_{LRG}$ is a
vector containing the galaxy overdensities $g=\delta n/\bar n$ in each
SDSS pixel, and ${\bf x}_{lens}$ consists of the two components of ${\bf
v}$ at each \WMAP pixel.  (We will suppress the frequency indices 
$(QQ)$, $(QV)$, etc. on ${\bf v}$ for clarity; it is understood that the 
analysis below is repeated for each pair of frequencies.)  The covariance 
of ${\bf x}$ is then:
\beq
{\sf C}\equiv
\langle {\bf xx}^T\rangle = \left( \begin{array}{cc} {\sf C}^{(LRG)} & 
{\sf C}^{(\times)} \\ {\sf C}^{(\times)T} & {\sf C}^{(lens)} 
\end{array} \right).
\label{eq:c}
\eeq
The cross-correlation matrix ${\sf C}^{(\times)}$ has components:
\beq
C^{(\times)}_{i,jK} = \sum_{lm} C_l^{gv} Y_{lm}^\ast(\nhat_i) {\bf 
Y}_{lm}^\parallel(\nhat_j)\cdot\hat{\bf e}_K,
\eeq
where $i$ represents an SDSS pixel index, $j$ is a \WMAP pixel index, and
$K=\hat\theta,\hat\phi$ indicates which component of the vector ${\bf v}$
is under consideration.  We bin the cross-spectrum $C_l^{gv}$ into bands,
\beq
C_l^{gv} = \sum_A c^A \left\{ \begin{array}{lcl} 1 & & l_{\rm min}(A)\le 
l<l_{\rm max}(A) \\
0 & & {\rm otherwise} \end{array} \right.,
\label{eq:crossbinned}
\eeq
and take the $c^A$ as the parameters to be estimated.

\begin{figure}
\includegraphics[angle=-90,width=3in]{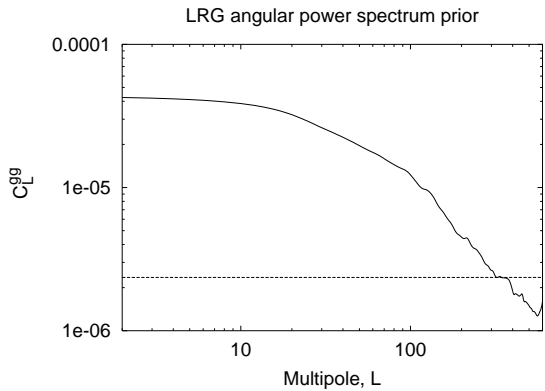}
\caption{\label{fig:lrgcl}The prior power spectrum used for the LRG map,
obtained by application of a pseudo-$C_l$ estimator to the SDSS scan
region.  The dashed line shows the Poisson noise.}
\end{figure}

In order to construct an optimal estimator for the galaxy-convergence
cross-spectrum, we need a prior auto-correlation matrix for the LRGs and
for the lensing map.  (This is a ``prior'' in the sense of quadratic
estimation theory \cite{1997MNRAS.289..285H, 1997MNRAS.289..295H,
1997PhRvD..55.5895T, 2003NewA....8..581P}, and has nothing to do with
Bayesian priors.)  We take a prior of the form
\beq
{\sf C}^{(LRG)}_{ij} = \sum_{lm} C_l^{gg} Y_{lm}^\ast(\nhat_i)
   Y_{lm}(\nhat_j) + N\delta_{ij},
\eeq
where $C_l^{gg}$ is the galaxy power spectrum (excluding Poisson noise)  
and $N$ is the noise variance per pixel.  We have taken $N$ to be the
reciprocal of the mean number of galaxies per pixel, appropriate for
Poisson noise (we use the mean galaxy density per pixel to avoid biases
associated with preferential weighting of pixels containing fewer
galaxies).  The prior power spectrum $C_l^{(LRG)}$ is determined by
application of a pseudo-$C_l$ estimator to the LRG maps; the resulting
power spectrum is shown in Fig.~\ref{fig:lrgcl}.  We have set 
$C_l^{gg}=0.01 \gg C_{l\ge 2}^{gg}$ for $l=0,1$ to reject the galaxy 
monopole (``integral constraint'') and dipole modes from the 
cross-correlation analysis.

It can be shown \cite{1997MNRAS.289..285H, 1997MNRAS.289..295H,
1997PhRvD..55.5895T} that for Gaussian data with small ${\sf
C}^{(\times)}$, the optimal estimator for the $c^A$ would be obtained by
taking the unbiased linear combinations of the ${\bf x}_{LRG}^T{\sf
C}^{(LRG)-1}{\sf P}_A{\sf C}^{(lens)-1}{\bf x}_{lens}$.  These 
estimators are frequently called ``QML'' or quadratic maximum-likelihood 
estimators, although they are, strictly speaking, not maximum-likelihood.
We do not have the full matrix ${\sf C}^{(lens)}$, and
our only knowledge of this matrix comes from simulations.  Thus we have 
instead constructed, for each lensing map ${\bf v}^{\alpha\beta}$, the 
quadratic combinations:
\beq
Q_A = {\bf x}^T_{LRG} {\sf C}^{(LRG)\,-1} {\partial {\sf 
C}^{(\times)}\over\partial c^A} {\bf x}_{lens},
\label{eq:qa}
\eeq
where $A$ represents a band index.  This differs from the QML estimators 
in that the ${\sf C}^{-1}$ weighting is applied only to the LRGs, while 
uniform weighting is applied to the CMB lensing map ${\bf v}$; thus 
Eq.~(\ref{eq:qa}) can be 
viewed as a sort of half-QML, half-pseudo-$C_l$ estimator for the 
cross-spectrum.  The expectation value of $Q_A$ can be
determined from Eq.~(\ref{eq:c}); it is:
\beq
\langle Q_A\rangle = \Tr \left[ {\partial{\sf C}^{(\times)T}\over 
\partial c^B} {\sf C}^{(LRG)\,-1} 
{\partial{\sf C}^{(\times)}\over\partial c^A} \right] c^B
\equiv R_{AB} c^B,
\label{eq:rab}
\eeq
which defines the response matrix $R_{AB}$.  Note that unlike the response
matrix of the optimal quadratic estimator, $R_{AB}$ is not equal to the
Fisher matrix.  The trace in Eq.~(\ref{eq:rab}) may be computed using a
stochastic-trace algorithm:
\beq
R_{AB} = \langle ({\sf C}^{(LRG)\,-1}{\bf y})^T
{\partial {\sf C}^{(\times)}\over\partial c^A}
{\partial {\sf C}^{(\times)T}\over\partial c^B} {\bf y}\rangle,
\label{eq:stotrace}
\eeq
where ${\bf y}$ is a random vector of length $N_{pix}^{(LRG)}$ consisting
of $\pm 1$ entries.  The vectors in Eqs.~(\ref{eq:qa}) and 
(\ref{eq:stotrace}) are constructed in pixel space.  Harmonic space is 
used only as an intermediate step in the convolutions required to compute 
the matrix-vector multiplications, e.g. ${\partial {\sf 
C}^{(\times)T}\over\partial c^B} {\bf y}$; these are computed by the usual 
method of converting to harmonic space, multiplying by $\partial 
C_l^{gv}/\partial c^B$, and converting back to real space.  The matrix 
inverse operations are performed iteratively as described in 
Appendix~\ref{app:prec}.  This method allows us to easily compute 
estimators for the band cross-powers,
\beq
\hat c^A = [R^{-1}]^{AB}Q_B.
\eeq
While this estimator is manifestly unbiased, we do not know its
uncertainty because we do not know the covariance ${\sf C}^{(lens)}$ of
the ${\bf v}$-field.  We determine the uncertainty via a Monte Carlo
method: we construct random CMB realizations according to the null 
hypothesis of no lensing in all eight DAs used for
the lensing reconstruction, and feed them through the lensing
reconstruction pipeline (Sec.~\ref{sec:lensing}).

In order to estimate the galaxy bias from the binned cross-power spectrum
estimators $\hat c^A$, we need to know the response of each estimator
$\hat c^A$ to the galaxy bias, $d\langle \hat c^A\rangle/db_g$.  This is 
given by
\begin{equation}
{d\langle \hat c^A\rangle \over db_g} = [R^{-1}]^{AB} \Tr\left[ {d{\sf 
C}^{(\times)T}\over db_g} {\sf C}^{(LRG)-1}
{\partial {\sf C}^{(\times)}\over\partial c^B} \right],
\label{eq:bgresponse}
\end{equation}
which is computed by a stochastic-trace algorithm analogous to 
Eq.~(\ref{eq:stotrace}).  The galaxy-${\bf v}$ correlation matrix 
$d{\sf C}^{(\times)T}/db_g$ is
\begin{equation}
{dC^{(\times)}_{i,jK}\over db_g} = \sum_{lm} R_l{dC^{g\kappa}\over db_g}
   Y_{lm}^\ast(\nhat_i)
   {\bf Y}^\parallel_{lm}(\nhat_j)\cdot\hat{\bf e}_K,
\end{equation}
where $R_l$ is the lensing response factor of Eq.~(\ref{eq:calib}).  If 
we knew $C^{\bf vv}$, it would be optimal to use $C^{{\bf vv}\,-1}$ 
weighting, in which case we could simply use $d{\sf C}^{(\times)T}/db_g$ 
as a cross-power template with no loss of information.  However since we 
have not calculated $C^{\bf v\bf v}$, and our only information on this 
covariance matrix comes from the ability to generate random realizations 
of ${\bf v}$, we cannot do this.

\subsection{Simulations}
\label{sec:sims}

Simulating lensed and unlensed CMB maps is necessary both for verifying
the analysis pipeline as well as for determining the optimal weighting of
Sec.~\ref{sec:falm}.  The general procedure used here is to generate a
simulated primary CMB temperature $T_{lm}$, convergence $\kappa_{lm}$, and
galaxy density fluctuation $\delta n_{lm}$ in harmonic space.  These are
Gaussian random fields and hence it is a straightforward matter to produce
random realizations from the power spectra and cross-spectra of $T$,
$\kappa$, and $\delta n$.  After generation of the realization, the
primary temperature and deflection field (generated from the convergence,
assuming an irrotational deflection field) are pixelized in HEALPix
resolution 10 (12,582,912 pixels of solid angle $11.8$ arcmin$^2$ each).  
The lensed CMB temperature $\tilde T$ is then computed in real space from
Eq. (\ref{eq:t-tilde}).  Because the ``deflected'' HEALPix pixels no
longer lie on curves of constant latitude, we use a non-isolatitude
spherical harmonic transform (see Appendix~\ref{app:nilsht}; we have used
parameters $L'=6144$ and $K=11$ since high accuracy is required) to
evaluate Eq. (\ref{eq:t-tilde}).  The beam convolution relevant to each DA
is then applied by converting to harmonic space, multiplying by
$B_l^\alpha$ and the pixel window function, and converting back to real
space.  Finally the simulated CMB temperature field $\dot T(\nhat)$ is
degraded to HEALPix resolution 9, and appropriate Gaussian ``instrument''
noise is added independently to each pixel.  Note that the resolution 10
pixels are used here to improve the fidelity of the simulation, in
particular to ensure that the effects of the elongated HEALPix pixels on
the lensing estimator are properly simulated.

A crude model for the \WMAP beam ellipticity is incorporated into the 
simulations as follows.  At each point, we have
\begin{equation}
\dot T^\alpha(\nhat) ={1\over 2}\sum_{{\rm side}=A,B} \sum_{lm\mu}
B_{l\mu} T_{lm} Y_{lm}^\mu(\nhat) \langle e^{i\mu\psi}\rangle,
\label{eq:td}
\end{equation}
where the $Y_{lm}^\mu$ are spin-weighted spherical harmonics and
the beam moments $B_{l\mu}$ are the multipole moments of the beam 
in instrument-fixed coordinates (with the ``North Pole'' along the 
boresight and the $\phi'=0$ meridian in the scan direction):
\begin{equation}
B_{l\mu} = \sqrt{4\pi\over 2l+1} \int_{4\pi} B(\nhat_{inst}) 
Y_{l\mu}^\ast(\nhat_{inst}) d^2\nhat_{inst}.
\end{equation}
The average value $\langle e^{i\mu\psi}\rangle$ is taken over the position 
angles of the instrument when $\nhat$ is scanned.  The sum over sides 
is over the two sides of \WMAPc.  Equation~(\ref{eq:td}) is only an 
approximation because (i) the two sides of the differencing assembly may 
not scan each pixel exactly the same number of times, or may have slightly 
different weights; and (ii) because \WMAP is a differential instrument, 
$\langle \dot T^\alpha(\theta,\phi) \rangle$ is also affected by 
beam-ellipticity effects on other parts of the sky.  Since the two beams 
of a given DA are separated by $\sim 140$ degrees, this results in an 
``echo'' of a given microwave source at a separation of 140 degrees 
\cite{2003ApJS..148...63H} (and higher-order echoes should also be 
present); we have neglected these.

We have made several further approximations to Eq.~(\ref{eq:td}) in order
to speed up the simulations.  First, we have only included the ellipticity
modes $\mu=\pm 2$, since these dominate the difference between the
azimuthally symmetrized beam and the true beam.  The beam ellipticity is
thus described by the real and imaginary parts of $B_{l,2}$; recall
$B_{l,-2}=B_{l,2}^\ast$.  We have calculated $B_{l,2}$ by taking the
non-isolatitude spherical harmonic transform of the \WMAP beam maps
\cite{2003ApJS..148...39P}. Secondly, because the side $A$ and $B$ beams
are approximate mirror-images of each other, we have only considered the
component of the beam ellipticity along the scan direction.  The component
of the beam ellipticity at 45 degrees to the scan direction is suppressed
because, to the extent that the side $A$ and $B$ beams are mirror images
and scan each pixel the same number of times, this component cancels in
Eq.~\ref{eq:td} when we sum over the two sides.

Finally, we have used a simple model for the scan pattern $\langle
e^{i\mu\psi}\rangle$ for each DA.  The \WMAP scan pattern is crudely 
approximated as a rotation around the spacecraft $-Z$ axis, followed by a 
precession of this axis in a 22.5 degree radius circle around the anti-Sun 
point, followed by rotation of the anti-Sun point along the ecliptic 
plane.  Relative to the spacecraft $-Z$ axis, the effective number of 
observations in one rotation is: $N_{obs}(\theta,\phi) = 
-N_{obs}\langle e^{2i\psi}\rangle = K 
\delta(\cos\theta-\cos\theta_c)/2\pi$, 
where $\theta_c$ is the angle between the instrument boresight and the 
spacecraft $-Z$ axis, and $K$ is a constant.  We can convert these to 
harmonic space in the spacecraft coordinates,
\begin{eqnarray}
[N_{obs}]_{lm}(-Z) &=& K \delta_{m0} Y_{l0}(\theta_c,0);
\nonumber \\
\left[N_{obs}\langle e^{2i\psi}\rangle \right]_{lm}(-Z) &=& -K \delta_{m0}
Y^2_{l0}(\theta_c,0).
\end{eqnarray}
Averaged over the precession cycle of \WMAPc, this becomes
\begin{eqnarray}
[N_{obs}]_{l0}({\rm av}) &=& K
P_l(\cos 22.5{\rm ~deg})P_l(0) Y_{l0}(\theta_c,0);
\nonumber \\
\left[ N_{obs}\langle e^{2i\psi}\rangle \right]_{l0}({\rm av}) &=& -K
P_l(\cos 22.5{\rm ~deg})
\nonumber \\ && \times
P_l(0) Y_{l0}^2(\theta_c,0).
\end{eqnarray}
(The $m\neq 0$ moments vanish.) Once these have been obtained, we may
transform back to real space and find $\langle e^{2i\psi}\rangle$ by
division.  This is then rotated from the ecliptic to the Galactic
coordinate system.  To speed up computation, the elliptical correction
to the beam was only computed on the HEALPix resolution 9 grid whereas
the dominant circular part was computed on the resolution 10 grid and
then degraded by pixel-averaging.

\section{Results}
\label{sec:results}

\subsection{Galaxy-convergence cross-spectrum}

\begin{figure*}
\includegraphics[angle=-90,width=6in]{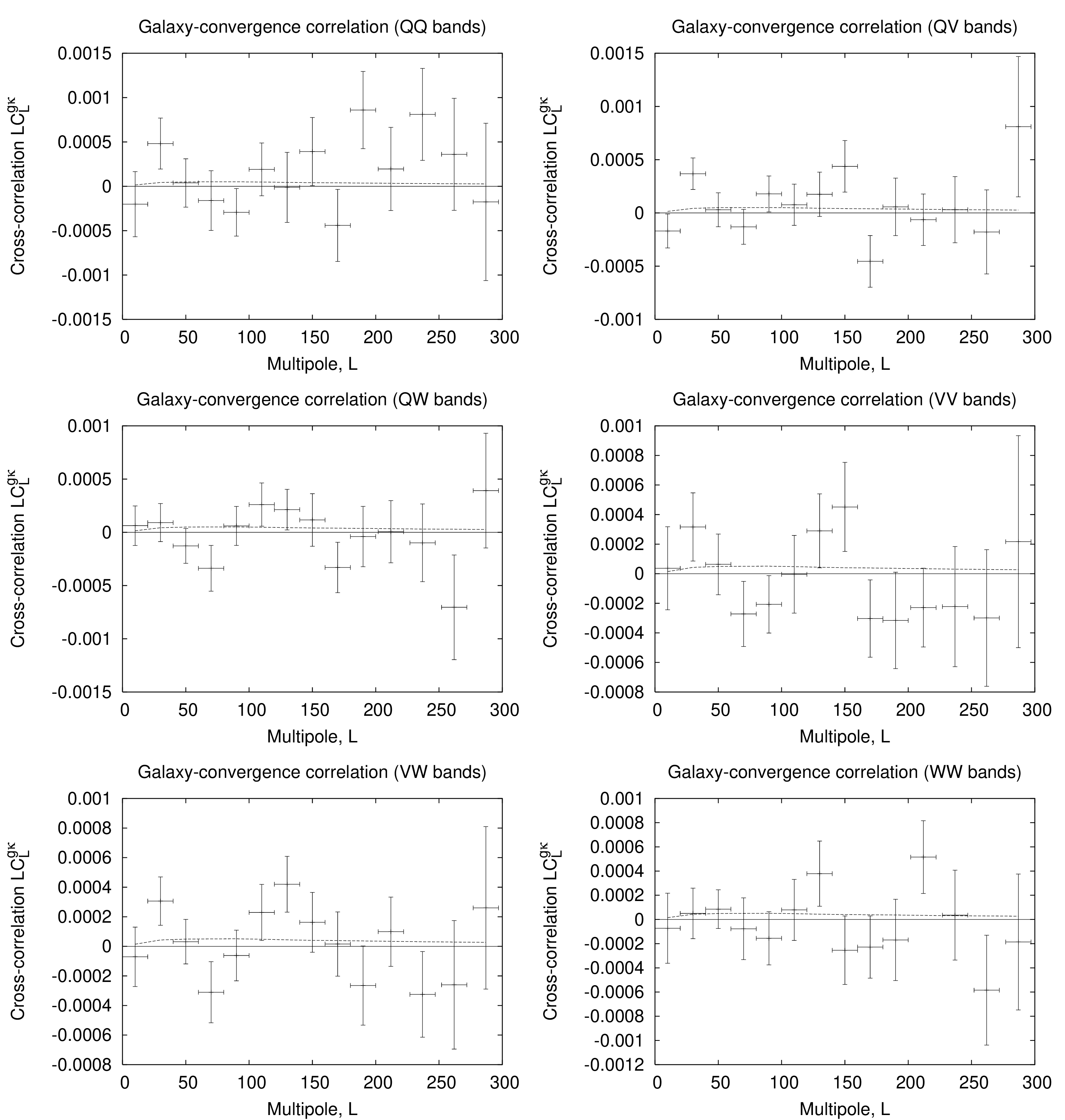}
\caption{\label{fig:cgk}The galaxy-convergence correlation using the 
Kp05\istenpst cut, for each of the six combinations of frequencies.  The 
error bars are strongly correlated across different frequencies.  The 
dashed curve shows the theoretical signal for our best-fit value of 
$b_g=1.81$.}
\end{figure*}

\begin{figure}
\includegraphics[angle=-90,width=3in]{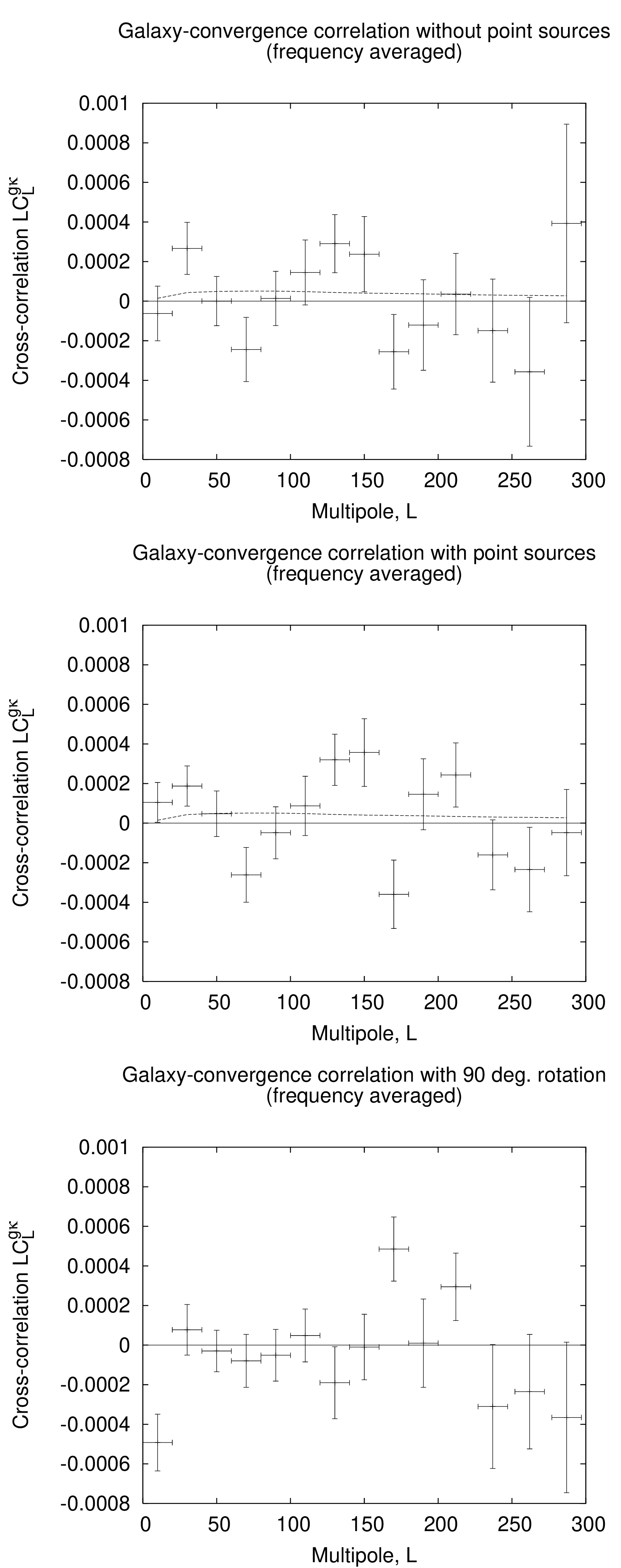}
\caption{\label{fig:ckgmulti}The galaxy-convergence correlation, using the 
frequency-averaged ${\bf v}^{(TT)}$ map.  The top panel shows the 
correlation using the Kp05\istenpst mask (which rejects point sources).  
The middle panel shows the correlation using the Kp05\isten mask, which 
does not reject point sources.  The bottom panel shows the correlation 
with the ${\bf v}^{(TT)}$ field rotated by 90 degrees; this should be 
zero in the absence of systematics (see Sec.~\ref{sec:ninety}).  The error 
bars are from simulations as described in Sec.~\ref{sec:amp}.  The dashed 
curve shows the theoretical signal for $b_g=1.81$.}
\end{figure}

The individual cross-spectra obtained at different frequencies are shown 
in Fig.~\ref{fig:cgk}.  The frequency-averaged cross-spectrum is shown in 
Fig.~\ref{fig:ckgmulti}, both with and without point sources.

\subsection{Amplitude determination}
\label{sec:amp}

We estimate the bias amplitude $b_g$ by fitting the observed
galaxy-convergence cross-spectrum $C^{g\kappa}_l$ to the theoretical
model, Eq.~(\ref{eq:limber}).  We begin by obtaining the covariance matrix
$\hat\Gamma$ of $\hat c^A$ as determined from $M=50$ simulations:
\beq
\hat\Gamma^{AB} = {1\over M-1} \sum_{i=1}^M
   (\hat c^A_i - \bar c^A)(\hat c^B_i - \bar c^B).
\label{eq:gammaest}
\eeq
(The simulated $\hat c^A$ are
generated by producing a random realization of the CMB as described in
Sec.~\ref{sec:sims} with no lensing, feeding it through the lensing
pipeline, and correlating it against the {\em real} SDSS LRG map.)
The bias is estimated from the weighted average of
the observed cross-powers in our $N=14$ bands:
\beq
\hat b_g = \frac{
\sum_{A=1}^N \hat c^A {dc^A_{th}\over db_g}/\hat\Gamma^{AA}} {
\sum_{A=1}^N \left({dc^A_{th}\over db_g}\right)^2/\hat\Gamma^{AA} },
\label{eq:hatbg}
\eeq
where $c^A_{th}$ is the theoretical prediction for the binned
cross-spectrum, Eq.~(\ref{eq:crossbinned}); this is directly proportional
to the bias, $c^A_{th}\propto b_g$.  The response $dc^A_{th}/db_g$ is
obtained from Eq.~(\ref{eq:bgresponse}). It is trivially seen that
Eq.~(\ref{eq:hatbg}) is an unbiased estimator of $b_g$, regardless of the
covariance of the $\hat c^A$.  Since we are working in harmonic space with
bands of width $\Delta l\ge 20>\Delta\theta^{-1}$, where $\Delta\theta$ is
the typical width of the survey region (in radians), the different
$l$-bands are very weakly correlated, so we have not attempted to further
optimize the relative weights of the various cross-power estimators $\hat
c^A$.

The most obvious way to estimate the uncertainty in $\hat b_g$ by noting 
that Eq.~(\ref{eq:hatbg}) is a linear function of the $c^A$, and 
substituting in the covariance matrix $\hat\Gamma^{AB}$ of the $\{c^A\}$:
\beq
\sigma(\hat b_g;{\rm ~incorrect}) = \frac{ \sum_{A,B=1}^N \hat\Gamma^{AB} 
{dc^A_{th}\over
db_g}{dc^B_{th}\over db_g} /\hat\Gamma^{AA}\hat\Gamma^{BB}}
{ \left[ \sum_{A=1}^N \left({dc^A_{th}\over db_g}\right)^2/\hat\Gamma^{AA} 
\right]^2}.
\label{eq:sigbgwrong}
\eeq
This calculation is incorrect for finite number $M<\infty$ of simulations
because it neglects the fact that the $\hat\Gamma^{AB}$ are themselves
random variables.  One approach to the problem is to take a sufficiently
large number of simulations $M$ that the error in
Eq.~(\ref{eq:sigbgwrong}) becomes negligible.  The difficulties in this
approach are that it could be very computationally intensive; we do not
know whether $M$ simulations are ``enough'' unless we try even larger
values of $M$ to check convergence.  An alternative method, which we have
used here, is to run an additional $M'=50$ simulated realizations of
$\{\hat c^A\}$ (identical to those used to compute $\hat\Gamma^{AB}$
except for the random number generator seed), compute the bias $\hat b_g$
from them, and then compute their sample variance.  The resulting error
bars can be analyzed using the well-known Student's $t$-distribution.  
The ``$1\sigma$'' $t$ error bars (which have 49 degrees of freedom)
obtained by this method are shown in Table~\ref{tab:bias}.  The mean bias
values obtained from these 50 random realizations are shown in the
``random'' column in the table. Also shown in Table~\ref{tab:bias} (in the
``foreground'' columns) are the results obtained by feeding the Galactic
foreground templates of Sec.~\ref{sec:gf-cmb} through the lensing pipeline
and correlating these with the real LRG map.

In the case of the Kp05\isten cut (last column in Table~\ref{tab:bias}),
which does not reject point sources, the ${\bf v}^{(QQ)}$, ${\bf
v}^{(QV)}$, and ${\bf v}^{(QW)}$ maps have power spectra that are boosted
significantly by point source contamination (see Sec.~\ref{sec:psfg}).  
Therefore, even if the correlation of the point sources with the galaxies
can be neglected, the error $\sigma(b_g)$ obtained in these bands for the
Kp05\isten mask is probably underestimated, as noted in the table.

\begin{table*}
\caption{\label{tab:bias}The bias $b_g$ estimated using several frequency 
ranges (first column).  The second column gives the bias $b_g$ obtained 
with the point sources removed (Kp05\istenpst cut); these are the numbers 
that should be thought of as our ``result.''  The uncertainties are 
$1\sigma$ with $t$ errors (49 d.o.f.); the first error given used the 
elliptical beam simulations, the error in parentheses is obtained from 
circular beam simulations (Sec.~\ref{sec:beam}).  The ``TT'' 
frequency combination is a weighted average of QV, QW, VV, VW, 
and WW.  The column labeled ``foreground'' show the bias $b_g$ obtained 
by correlating the ${\bf v}$ maps derived from the Galactic foreground 
maps of Sec.~\ref{sec:gf-cmb} against the LRG map.  Similarly, the column
labeled ``random'' show the bias $b_g$ obtained using the average of the 
50 random realizations of ${\bf v}$ in place of the \WMAPc-derived ${\bf 
v}$ map.  The column labeled ``TT weight'' shows the bias determined by 
using the same weights as a function of $l$ as for the TT frequency 
combination in Eq.~(\ref{eq:hatbg}); of course this has no effect on 
$b_g(TT)$.
The final column, labeled ``Kp05\isten'', is the bias obtained without the 
point source cut.}
\begin{tabular}{cdcdcdcdcd}
\hline\hline
Frequency
 & \mbox{Bias, $b_g$} \backtwen
& & \mbox{foreground} \backtwen
& & \mbox{random} \backtwen
& & \mbox{TT weight} \backtwen
& & \mbox{$b_g$, Kp05\isten} \backtwen
\\
\hline
QQ  & +3.62\pm 4.48 (\pm 4.33) \backtwen
&\fwten & -0.001640 & & -0.34\pm 0.63 & & +3.13\pm 4.55
& & +6.30\pm 3.93\footnotemark[1] \\
QV  & +3.10\pm 2.19 (\pm 2.00) \backtwen
&\fwten & -0.000001 & & +0.07\pm 0.31 & & +3.39\pm 2.29
& & +3.93\pm 1.99\footnotemark[1] \\
QW  & -0.11\pm 2.53 (\pm 2.35) \backtwen
&\fwten & -0.000480 & & +0.06\pm 0.36 & & -0.20\pm 2.56
& & +0.48\pm 2.43\footnotemark[1] \\
VV  & -1.41\pm 2.95 (\pm 3.13) \backtwen
&\fwten & -0.000737 & & +0.88\pm 0.42 & & +0.34\pm 3.04
& & +0.58\pm 2.35 \\
VW  & +2.63\pm 2.56 (\pm 2.11) \backtwen
&\fwten & -0.000617 & & +0.35\pm 0.36 & & +2.58\pm 2.62
& & +2.51\pm 2.03 \\
WW  & +0.23\pm 3.11 (\pm 2.71) \backtwen
&\fwten & -0.000817 & & +0.20\pm 0.44 & & +0.65\pm 3.22
& & -0.75\pm 2.63 \\
TT  & +1.81\pm 1.92 (\pm 1.72) \backtwen
&\fwten & -0.000449 & & +0.23\pm 0.27 & & +1.81\pm 1.92
& & +2.43\pm 1.58\footnotemark[1] \\
\hline\hline
\end{tabular}
\footnotetext[1]{Because of point sources, the ${\bf v}$ maps from these 
bands contain excess power in the Kp05\isten region.  Thus the 
simulation error bars shown here are likely underestimates.}
\end{table*}

The $\hat\chi^2$ values for fits to zero signal are shown in 
Table~\ref{tab:chi2}.  These are obtained using the 14 band cross-power 
spectra (Fig.~\ref{fig:ckgmulti}), and the $14\times 14$ covariance matrix 
is obtained from 100 simulations,
\begin{equation}
\hat\chi^2 = [\hat\Gamma^{-1}(100{\rm ~sims})]_{AB}c^Ac^B.
\end{equation}
Because the number of simulations is finite, there remains some noise in
this covariance matrix and this must be taken into account in interpreting
the $\hat\chi^2$.  In particular, the $\hat\chi^2$ variable does not
exactly follow the standard $\chi^2$ (so we have denoted it with a hat).  
The distribution and $p$-values can, however be calculated as described in 
Ref.~\cite{2004astro.ph..3255H}, Appendix D.  As noted previously, the 
errors for the Kp05\istenpst mask in the QQ, QV, and QW combinations are 
suspect.

\begin{table}
\caption{\label{tab:chi2}The $\hat\chi^2$ values obtained for fits to zero 
signal from the galaxy-convergence cross-spectrum.  The first column 
shows the results with the Kp05\isten mask; the second using the 
Kp05\istenpst mask; and the third using the Kp05\istenpst mask with 90 
degree rotation of ${\bf v}$.  The $\chi^2$ has 14 degrees of freedom 
(the 14 $l$-bins shown in Fig.~\ref{fig:ckgmulti}) and the covariance 
matrices were obtained from the 100 simulations described in 
Sec.~\ref{sec:amp}.  As described in the text, the finite number of 
simulations means that the expectation value of the $\hat\chi^2$ is not 14 
but is larger due to uncertainty in the covariance matrix; the mean of 
the $\hat\chi^2$ distribution is 16.5 and the standard deviation is 
6.8.  We have also given the cumulative probability distributions.}
\begin{tabular}{cccdccdccd}
\hline\hline
Freq. & & \multicolumn{2}{c}{\mbox{Kp05\isten}} & &
\multicolumn{2}{c}{\mbox{Kp05\istenpst}} & &
\multicolumn{2}{c}{\mbox{Kp05\istenpst}} \\
 & & & & & & & & \multicolumn{2}{c}{\mbox{+90 degrees}} \\
 & & $\hat\chi^2$ & P(<\hat\chi^2)\backten 
 & & $\hat\chi^2$ &
P(<\hat\chi^2)\backten \!\!
 & & $\hat\chi^2$ & P(<\hat\chi^2)\backten \!\! \\
\hline
QQ & & 51.51\footnotemark[1] & \backten 0.9996\footnotemark[1]
\backten
& & 16.24 & 0.55  & & 28.24 & 0.940 \\
QV & & 37.01\footnotemark[1] & \backten 0.990\footnotemark[1]
\backten
& & 27.51 & 0.931 & & 21.89 & 0.81  \\
QW & & 16.74\footnotemark[1] & \backten 0.58\footnotemark[1]
\backten
& & 11.74 & 0.26  & & 21.58 & 0.80  \\
VV & & 19.08 & \backten 0.70   \backten & & 11.86 
& 0.27  & & 24.85 & 0.89  \\
VW & & 25.30 & \backten 0.90   \backten & & 20.28 
& 0.75  & & 20.06 & 0.74  \\
WW & & 11.54 & \backten 0.25   \backten & & 11.36 
& 0.24  & & 21.82 & 0.81  \\
TT & & 26.90\footnotemark[1] & \backten 0.922\footnotemark[1]
\backten
& & 19.97 & 0.74  & & 30.71 & 0.963 \\
\hline\hline
\end{tabular}
\footnotetext[1]{Because of point sources, the ${\bf v}$ maps from these 
bands contain excess power in the Kp05\isten region.  Thus the 
simulation error covariance matrices are likely underestimates, and hence 
the $\hat\chi^2$ and $P(<\hat\chi^2)$ values are suspect.}
\end{table}

\section{Systematic errors}
\label{sec:systematics}

\subsection{Ninety-degree rotation test}
\label{sec:ninety}

One of the standard systematics tests in weak lensing studies using
galaxies as sources has been to rotate all of the galaxies by 45 degrees
and look for a shear signal.  The 45 degree rotation is used because it
interconverts $E$ and $B$ modes, and in the absence of systematics there
should be no $B$-mode signal.  In the case of CMB lensing using the vector
estimator ${\bf v}$, the analogous test is to rotate ${\bf v}$ by 90
degrees (thereby interchanging the longitudinal and transverse parity
modes).  This rotated map can be fed through the cross-correlation
pipeline in place of the original ${\bf v}$.  In the absence of
systematics, this gives zero signal; the error bars need not be the same
as for the longitudinal modes, but they can still be determined from
simulations as described in Sec.~\ref{sec:amp}.  The cross-spectrum is
shown in Fig.~\ref{fig:ckgmulti} and the $\hat\chi^2$ values in
Table~\ref{tab:chi2}.

The lowest-$l$ point in the rotated cross-spectrum
(Fig.~\ref{fig:ckgmulti}) is $3.4\sigma$ negative.  It is difficult to
assess the significance of this anomaly since it is an {\em a posteriori}
detection ($p=0.00127$ for the two-tailed $t$-distribution); in any case,
it is responsible for the relatively high $\hat\chi^2$ value ($p=0.037$)  
in the Kp05\istenpst +90~degree column of Table~\ref{tab:chi2}.  It is
unlikely that this correlation $\langle g^\ast v^{(TT)\perp}_{lm}\rangle$
represents any real astrophysical or cosmological effect, since it
violates parity.  This anomaly is also distinct from the much-discussed
``low quadrupole'' observed by \WMAPc, since the former is based on a
high-pass filtered CMB map with power coming predominately from CMB modes
with $l\sim$few$\times 10^2$. Another possible explanation would be some
source of excess power in the ${\bf v}$ map at low $l$, which would
increase the error bar relative to simulations and thus lower the
statistical significance of this point.  However if we take the ${\bf v}$ 
maps, and compute the un-deconvolved power spectrum
\beq
P = \sum_{l<20}\sum_{m=-l}^l \left| \sum_{\nhat\in{\rm Kp05}\cap{\rm S10}
   \setminus{\rm ps}_2} {\bf v}(\nhat)\cdot Y_{lm}^{\perp\ast}(\nhat)
   \right|^2
\eeq
for both the real ${\bf v}$ map and the 100 simulated maps, we find that
the real map has the 28th highest value of $P$ out of 101 maps, i.e.  
there is no evidence for excess power.  If this point is due to some
systematic, it must be present at all three frequencies, since this point
is negative by at least $1\sigma$ in all of the frequency combinations
except QQ, where the binned $C_l^{g\kappa}$ from the rotated map at $l<20$
is $(1.0\pm 3.5)\times 10^{-5}$.

It is thus difficult to explain the lowest-$l$ point in
Fig.~\ref{fig:ckgmulti} point in terms of any systematic error.  The true
test for whether this is in fact just a statistical fluctuation is to wait
for the error bars to become smaller with future \WMAP data and see
whether this point becomes more significant or goes away, and in the
former case whether it exhibits a frequency dependence.

\subsection{End-to-end simulation}
\label{sec:calib}

Another important systematic test is to verify, in an ``end-to-end''
simulation, that the lensing estimator and $C^{g\kappa}_l$ cross-spectrum
estimator are calibrated properly.  This can be done as follows. We run 50
simulations in which simulated Gaussian $g$ and $\kappa$ maps are
generated with the cross-spectrum appropriate for $b_g=1$.  The $\kappa$
map has the power spectrum $C^{\kappa\kappa}_l$ expected for a
$\Lambda$CDM cosmology, while the $g$ map is constructed from
$g_{lm}=C^{g\kappa}_l\kappa_{lm}/C^{\kappa\kappa}_l$.  In principle one
could add additional noise to $g$ to boost its power spectrum to match the
observed $C^{gg}_l$, but there is no reason to do this as it
increases the number of simulations required and has no effect on the
calibration.  The $\kappa$ maps are then used to generate lensed CMB maps
using the simulation code described in Sec.~\ref{sec:sims}, and the output
temperature maps $\dot T_{Q1}...\dot T_{W4}$ fed through the lensing
reconstruction pipeline and then the $C^{g\kappa}_l$ estimator.  Finally,
we estimate the bias in each simulation using Eq.~(\ref{eq:hatbg}).  This
output $b_g$ is the calibration factor $1+\zeta$ appropriate for
cross-correlation studies.

The calibration factors obtained from this procedure are shown in
Table~\ref{tab:e2e}; the table reveals that the lensing pipeline is
calibrated at the $\sim 20\%$ level.  Calibration factors of this order
have been observed in previous simulations \cite{2003PhRvD..67d3001H,
2004astro.ph..3075A} and have been investigated analytically
\cite{2003NewA....8..231C, 2003PhRvD..67d3001H, 2003PhRvD..67l3507K,
2003PhRvD..68h3002H}, where the main effect has been the non-linear
lensing effects (i.e. the order $\kappa^2$ and higher terms that have been
dropped in the Taylor expansion, Eq.~\ref{eq:tcov}).  In our case, the
calibration error may also have a contribution from the elliptical beam.  
In any case, the calibration errors are not significant at the level of
the current data (i.e. no detection).

\begin{table}
\caption{\label{tab:e2e}The calibration factors obtained via end-to-end 
simulations, for the Kp05\istenpst mask and various combinations of 
frequencies.  Error bars are $1\sigma$, $t$-distributed with 49 degrees 
of freedom.}
\begin{tabular}{ccc}
\hline\hline
Frequencies & & Calibration factor, $1+\zeta$ \\
\hline
QQ & & $1.17\pm 0.13$ \\
QV & & $1.24\pm 0.09$ \\
QW & & $1.19\pm 0.07$ \\
VV & & $1.31\pm 0.11$ \\
VW & & $1.15\pm 0.07$ \\
WW & & $0.92\pm 0.10$ \\
TT & & $1.18\pm 0.07$ \\
\hline\hline
\end{tabular}
\end{table}

\subsection{Beam effects}
\label{sec:beam}

We have used a crude model for the \WMAP beam ellipticity.  An incorrect
model for the beam can have three effects on the galaxy-convergence cross
spectrum and hence on the bias determination: it can (i) produce a shear
calibration bias in the ${\bf v}$ map (which may depend on the wavenumber
$l$ and orientation of the convergence mode in question, and may vary
across the sky as the effective beam varies); (ii) modify the noise
covariance matrix of ${\bf v}$; and (iii) introduce artifacts (i.e.  
biases) in the ${\bf v}$ map because it invalidates the assumption that
the signal is statistically isotropic.  The calibration problem is
obviously of concern for attempts to do precision cosmology with lensing.
However since we do not have a detection, the only effect of the
calibration bias is to affect our upper limits on the lensing signal.  
The change in the noise covariance of the lensing map ${\bf v}$ is
potentially more serious because it can alter the variance of our
cross-spectrum estimator and hence affect the statistical significance of
any lensing detection. Because artifacts in the ${\bf v}$ map do not
correlate with the galaxy distribution, they are essentially also a source
of spurious power, and for cross-correlation measurements they are only a
concern if their power spectrum is comparable to that of the noise.

The effect of the beam on the noise covariance can be addressed as
follows.  We have re-computed the uncertainties $\sigma(b_g)$ (see
Sec.~\ref{sec:amp}) using 50 simulations with a circular beam instead of
our elliptical beam model.  The uncertainties are shown in parentheses in
Table~\ref{tab:bias}.  They are at most 20\% different from the error
estimates obtained from the elliptical beams (but this may not be
significant because the $\sigma(b_g)$ values from simulations are
themselves drawn from a random distribution -- namely, the square root of
a $\chi^2$ distribution with 49 degrees of freedom -- and hence have an
uncertainty of $1/\sqrt{2\cdot 49}\approx 10$\%).  Since replacing our
model of the beam ellipticity with the inferior model of a circular beam
has only a $<20$\% effect on $\sigma(b_g)$, it is doubtful that
$\sigma(b_g)$ would be altered by more than this by use of an improved
beam model.

Finally, we come to the issue of the calibration.  Our estimator for the 
bias was constructed assuming a circular beam, and given that the true 
beam is not circular, we expect that it may be mis-calibrated, i.e. 
$\langle\hat b_g\rangle = (1+\zeta)\hat b_g$, where $\zeta$ is the 
calibration bias.  At present, the best way to test for such a bias is via 
simulations, such as those of Sec.~\ref{sec:calib}.  There we found a 
calibration bias of $\zeta = 0.18\pm 0.07$, which is not important at the 
level of the present data.

\subsection{Extragalactic foregrounds}
\label{sec:extra}

The lensing estimator of Eq.~(\ref{eq:vdef}) will respond not only to real
lensing signals, but to any other perturbations of the CMB.  Of greatest
concern is the contamination from extragalactic foregrounds, which may
induce spurious correlation of the lensing estimator with the galaxy
distribution since the extragalactic foregrounds (tSZ and point sources)
are expected to correlate with large scale structure.  The presence of the
extragalactic foregrounds causes the observed temperature $\hat T(\nhat)$
to be incremented by some amount $\Delta T^\alpha(\nhat)$.  Assuming that
the extragalactic foregrounds are not correlated with the primary CMB or
with instrument noise, this causes the expectation value of ${\bf v}$
(averaged over primary CMB and noise realizations) to be incremented by
(compare to Eq.~\ref{eq:harmonicv})
\beqa
\Delta \langle v^{\alpha\beta(\parallel)}_{lm} \rangle &=& (-1)^m 
\sum_{l'l''} {\cal K}_{ll'l''} \sum_{m'm''}
\threej{l}{l'}{l''}{-m}{m'}{m''}
\nonumber \\ && \times
{\Delta T^\alpha_{l'm'} \Delta T^\beta_{l''m''} + \Delta T^\beta_{l'm'} 
\Delta T^\alpha_{l''m''} \over 2}.
\label{eq:dq}
\eeqa
Thus the possible source of contamination of the convergence-galaxy
correlation signal $C^{g\bf v}_l$ is the correlation of the LRGs with {\em
quadratic} combinations of the foreground temperature.  In the case of the
tSZ foreground, the contribution to Eq.~(\ref{eq:dq}) can be broken up
into a ``single-halo'' term in which the two factors of $\Delta T$ come
from the same halo, and a ``two-halo'' term in which the two factors of
$\Delta T$ come from different halos.  The single-halo term exists even if
the tSZ halos are Poisson-distributed, whereas the two-halo term acquires
a nonzero value only from clustering of the halos.  Much of this section
will be devoted to an investigation of the properties of the quadratic
combinations in Eq.~(\ref{eq:dq}) and an assessment of their magnitude.  
Unfortunately, we will see that this does not result in useful constraints
on the foreground contamination to our measurement of $b_g$.

\subsubsection{Point sources}
\label{sec:psfg}

\begin{figure*}
\includegraphics[angle=-90,width=5.5in]{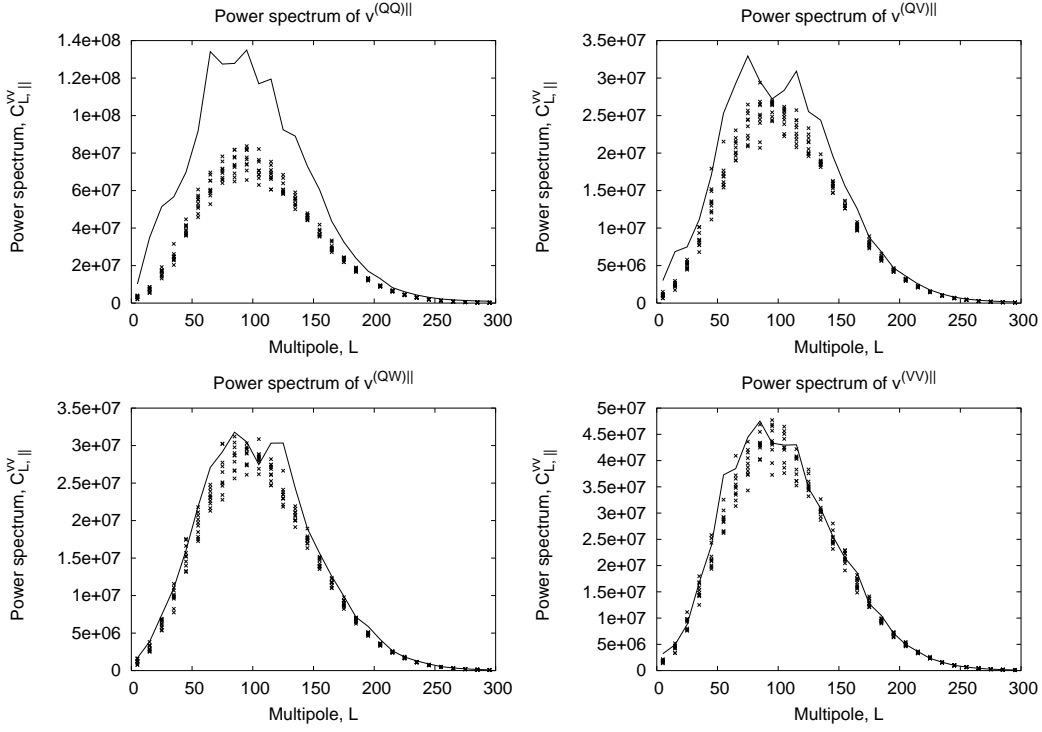}
\caption{\label{fig:vpowerwithps}The cut-sky power spectrum of the
longitudinal mode of ${\bf v}$, $C^{vv}_{l,\parallel}$; the power spectrum
of the actual map is shown as the solid line, while the points are from 10
simulations.  This figure is identical to Fig.~\ref{fig:vpower}, except
that we have used the Kp05\isten cut (i.e. in this figure the point
sources are not masked), and we have not shown the VW or WW spectra 
(which do not differ significantly from Kp05\istenps).  Because of 
contamination by point sources, the power in the QQ and QV maps is greater 
than in the simulations or the point source-cut maps.}
\end{figure*}

It is readily seen that point sources are a major contribution to the
power spectrum of ${\bf v}$, especially in the lower-frequency bands.  
This can be seen from Fig.~\ref{fig:vpowerwithps}, in which the power
spectrum $C^{vv}_{l,\parallel}$ is shown in the Kp05\isten region (in
which point sources are not masked).  The ${\bf v}^{(QQ)}$, ${\bf
v}^{(QV)}$, and ${\bf v}^{(QW)}$ maps are heavily contaminated while for
the higher-frequency maps point sources are subdominant to CMB
fluctuations.  While the contribution to the ${\bf v}$ autopower in
Kp05\isten is large, we are interested here in whether -- and at what
frequencies -- the point source contribution to ${\bf v}^{\alpha\beta}$
correlates with the LRG map when the Kp05\istenpst mask (which masks point
sources) is used.

\begin{figure}
\includegraphics[angle=-90,width=3in]{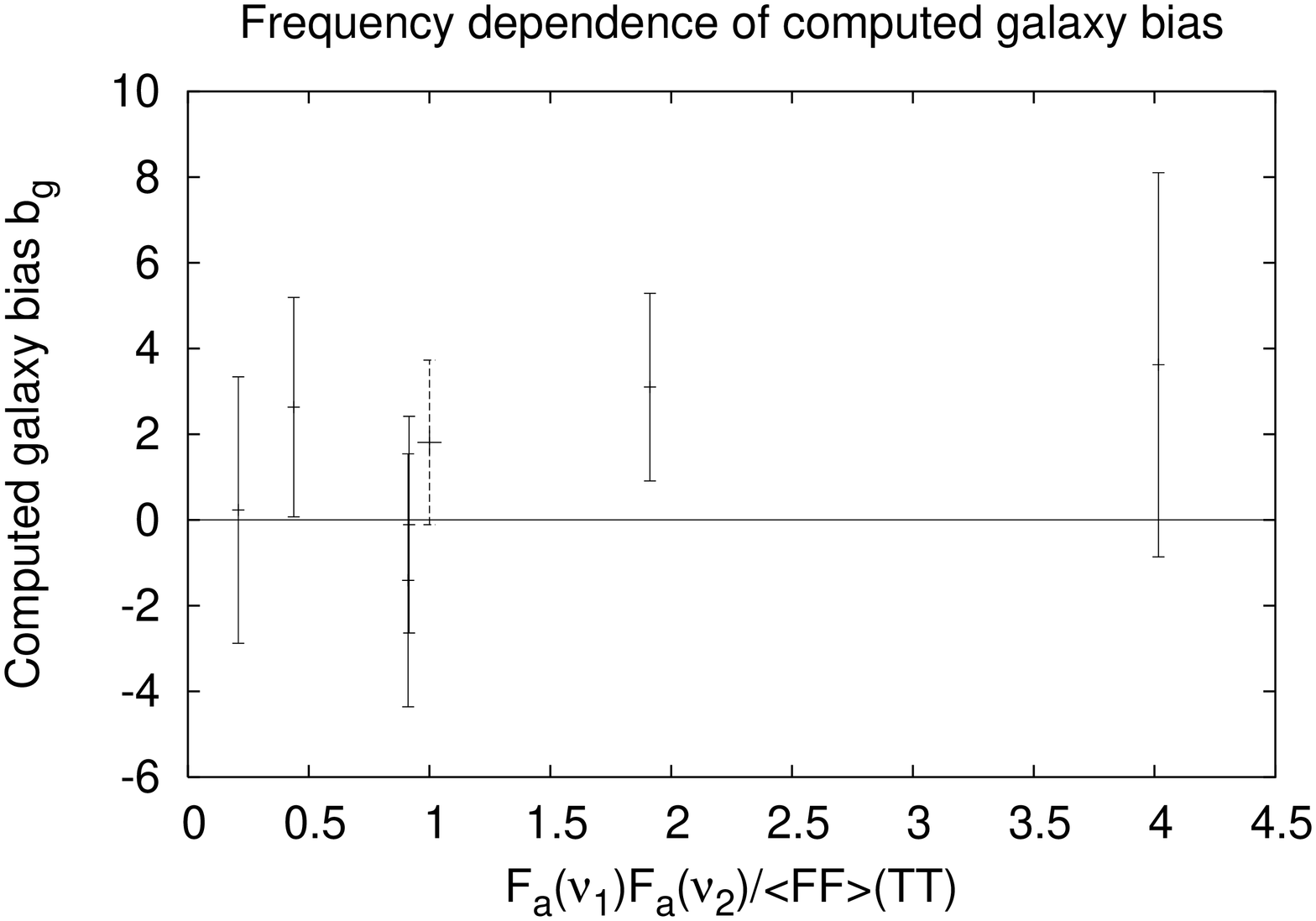}
\caption{\label{fig:biasfreq}The dependence of the computed galaxy bias 
$b_g$ on frequency.  The horizontal axis is $F_a(\nu_1)F_a(\nu_2)/0.249$, 
computed for a typical synchrotron spectrum ($\alpha=0.0$); the 
normalization is chosen so that the frequency-averaged $TT$ result has 
$F_a(\nu_1)F_a(\nu_2)/0.249=1$.  The horizontal line is zero, and the 
point with dashed error bars is $TT$.  (This point is of course the 
average of the other data points and contains no additional information.)  
The error bars are correlated; the $\chi^2$ for a frequency-independent 
bias is $6.31$ with 5 degrees of freedom.}
\end{figure}

A single point source with frequency-dependent flux $F_a(\nu)$ (in units
of blackbody $\mu$K~sr) at position $\nhat_a$ will produce a spurious
contribution to the temperature of
\beq
\Delta T^\alpha_{lm} = F_a(\nu^\alpha) Y_{lm}^\ast(\nhat_a).
\label{eq:psdt}
\eeq
Plugging this into Eq.~(\ref{eq:dq}), we find that the shift in 
$\langle{\bf v}\rangle$ is:
\beq
\Delta \langle v^{\alpha\beta(\parallel)}_{lm} \rangle = F_a(\alpha) 
F_a(\beta) Y_{lm}^\ast(\nhat_a) r_{PS}(l),
\label{eq:psdv}
\eeq
where the response function $r_{PS}(l)$ is:
\beq
r_{PS}(l) = \sqrt{(2l'+1)(2l''+1)\over 4\pi(2l+1)} 
\threej{l}{l'}{l''}{0}{0}{0} {\cal K}_{ll'l''}.
\eeq
In Table~\ref{tab:fa}, we show the product $F_a(\nu_1) F_a(\nu_2)$ for
several possible point source spectra.  Note that for the steep spectra
characteristic of \WMAP point sources ($\alpha\sim 0.0$), the
contamination of ${\bf v}^{(QQ)}$ and ${\bf v}^{(QV)}$ is far greater than
contamination of the higher-frequency lensing maps.  Therefore these
lower-frequency bands are a useful test of point source contamination of
the galaxy-convergence correlation.  The dependence of the estimated $b_g$ 
on the combination of frequencies, $F_a(\nu_1) F_a(\nu_2)$, is shown in 
Fig.~\ref{fig:biasfreq}.  If there were point source contamination of our 
measurement with spectral index $\alpha=0.0$, the points in 
Fig.~\ref{fig:biasfreq} would be expected to fall roughly along a line 
$b_g\propto F_a(\nu_1)F_a(\nu_2)$; this is only rough because 
the weighting of different $l$ bins is slightly different at different 
frequencies in Eq.~(\ref{eq:hatbg}), and the contaminating signal 
need not have the same angular dependence as the galaxy-convergence 
correlation.  If we re-calculate the six frequency combinations 
$b_g(QQ)...b_g(WW)$ using the same weighting of different $l$ bins as for 
the $b_g(TT)$ measurement, we get the values in the column in 
Table~\ref{tab:bias} labeled ``TT weight.''  Assuming a synchrotron-like 
spectrum for the point sources, a correlated least-squares fit of the form
\beq
b_g(\nu_1\nu_2) = b_g^{(0)} + \Delta b_g^{(PS)}(TT)
\frac{ F_a(\nu_1)F_a(\nu_2) }{ \langle F_aF_a\rangle(TT) }
\eeq
to the various frequency combinations will return for $b_g^{(0)}$ a
point-source-marginalized measurement of the bias, and for $\Delta
b_g^{(PS)}(TT)$ a measurement of the point-source contamination to the
unmarginalized $b_g(TT)$.  The results of such a fit are $b_g^{(0)} =
0.58\pm 2.36$ and $\Delta b_g^{(PS)}(TT) = 0.73\pm 1.18$; the two
measurements are of course anti-correlated with correlation coefficient
$\rho=-0.63$.  There is thus no evidence for point source contamination,
although the statistical errors are too large to definitively say whether
such contamination is present at the level of the signal.  It would be
useful to have lower-frequency information here in order to improve the
constraints, however this is not possible as the CMB multipoles used in
our analysis ($l\sim 350$) are not resolved by \WMAP K- and Ka-band
differencing assemblies.

\begin{figure}
\includegraphics[angle=-90,width=3in]{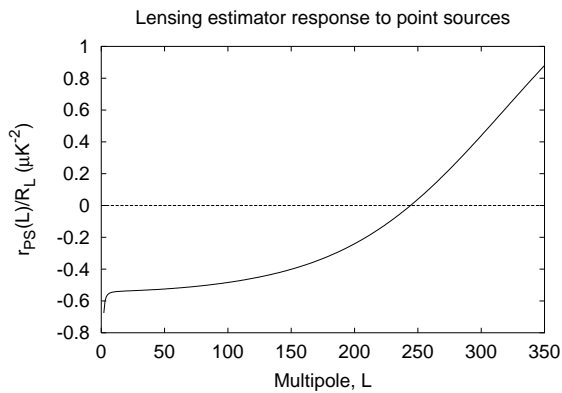}
\caption{\label{fig:ql}The ratio $r_{PS}/R_l$ describing the response of 
the lensing estimator ${\bf v}$ to contamination from 
Poisson-distributed point sources.  Note that on large scales this 
is negative, i.e. regions with more point sources are interpreted by the 
lensing estimator as regions of negative convergence (underdense 
regions).  At high $l$ the situation is reversed.}
\end{figure}

\begin{figure}
\includegraphics[angle=-90,width=3in]{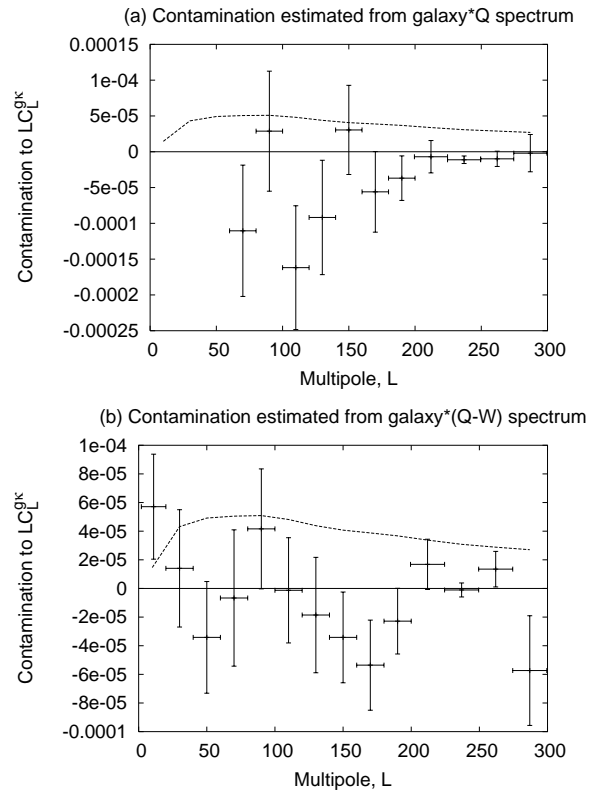}
\caption{\label{fig:pscl}(a) The contamination to the galaxy-convergence 
cross-spectrum due to point sources, based on the Q-band 
galaxy-temperature cross-spectrum in the Kp0\isten region (see text).  The 
dashed line is the 
best-fit signal from Sec.~\ref{sec:amp}, with $b_g=1.81$.  We have 
removed the $l<60$ points since these may contain ISW signal.  (b) The 
same figure, except derived from the difference of galaxy-temperature 
cross spectra in Q and W bands.  This cancels any contribution from ISW 
and reduces cosmic-variance errors at low $l$.}
\end{figure}

The contamination in $C^{g\kappa}_l$ from a point source can also be 
estimated from angular information since the point sources have an 
angular dependence (roughly Poisson) unlike that of the CMB.  
Equations~(\ref{eq:psdt}) and (\ref{eq:psdv}) give
\beqa
\Delta C^{g\kappa(TT)}_l &=& {r_{PS}(l)\over R_l} 
\left[ \sum_{\alpha\beta} 
a^{(TT)}_{\alpha\beta} {F(\alpha)F(\beta)\over F(Q)^2} \right]
\nonumber \\ && \times
F(Q) \Delta C^{gT(Q)}_l,
\label{eq:psco}
\eeqa
where $\Delta C^{gT(Q)}_l$ is the point source-induced galaxy-temperature 
cross spectrum in Q-band.  If take a typical spectrum $\alpha=0.0$ and 
a flux $F(Q)=1$~Jy (i.e. the brightest point sources not excluded by the 
\WMAP point source mask \cite{2003ApJS..148...97B, 2003ApJS..148..119K}), 
we find
\beq
\left[ \sum_{\alpha\beta}
a^{(TT)}_{\alpha\beta} {F(\alpha)F(\beta)\over F(Q)^2} \right]
F(Q) = 6.7\times 10^{-3} \mu{\rm K~sr}.
\eeq
The ratio $r_{PS}(l)/R_l$ is plotted in Fig.~\ref{fig:ql}. Of course, not
all of the point sources have $F(Q)=1$~Jy, but this is the worst-case
scenario since $\Delta C_l^{g\kappa}\propto F\Delta C_l^{gT}$, hence if
the galaxy-temperature cross-spectrum is coming from fainter sources the
contamination will be even less. (This scaling with $F$ occurs because the
spurious contribution to the galaxy-convergence cross spectrum is
quadratic in the flux whereas the contribution to the galaxy-temperature
cross spectrum is linear.)  We have computed the Q-band galaxy-temperature 
cross-spectrum using a QML estimator \cite{isw} on the Kp0\isten cut; if 
we take this cross-spectrum, and assume that at $l>60$ the 
cross-spectrum is entirely due to point sources with the synchrotron 
spectrum, the derived contamination to the galaxy-convergence spectrum is 
as shown in Fig.~\ref{fig:pscl}(a).  One can also use the difference 
between Q and W-band cross-spectra; in this case, if a synchrotron 
spectrum with $\alpha=0.0$ for the point sources is assumed, the 
difference $C_l^{gT}(Q)-C_l^{gT}(W)$ must be multiplied by 1.295 to 
recover the $C_l^{gT}(Q)$ used in Eq.~(\ref{eq:psco}).  This result is 
shown in Fig.~\ref{fig:pscl}(b); the error bars (obtained from 
simulations) are now smaller at low $l$ because the CMB fluctuations are 
suppressed.  The point-source induced error in the bias can be computed 
from the data in Fig.~\ref{fig:pscl}(b) by plugging these $\Delta 
C_l^{g\kappa}$ values into Eq.~(\ref{eq:hatbg}); the result is $\Delta 
b_g^{(PS)} = -0.14 \pm 0.51$.

We conclude that the point-source contamination to $C^{g\kappa}_l$ is at
most of the same order as the signal in this range of multipoles.  If one
ignores correlations between distinct point sources so that
Eq.~(\ref{eq:psco}) is valid and assumes the $\alpha=0.0$ spectrum, then
Fig.~\ref{fig:pscl}(b) suggests that the point source contamination is
less than the observed signal.

\begin{table}
\caption{\label{tab:fa}The point source frequency-dependent response
$F_a(\nu_1) F_a(\nu_2)$ for several point source spectra.  Included are
three power-law spectra, $T_{antenna}\propto \nu^{\alpha-2}$
($\alpha=-1.0$, $0.0$, and $+0.5$) and a nonrelativistic tSZ spectrum. We
have normalized to unit response in ${\bf v}^{(QQ)}$. The ``TT'' frequency 
combination is an average over the other pairs of frequencies weighted by 
$a^{(TT)}_\alpha\beta$.}
\begin{tabular}{ccccccccc}
\hline\hline
Bands & & \multicolumn{7}{c}{\mbox{$F_a(\nu_1)F_a(\nu_2)$}} \\
 & & $\alpha=-1.0$ & & $\alpha=0.0$ & & $\alpha=+0.5$ & & tSZ \\
\hline
QQ & & 1.000 & & 1.000 & & 1.000 & & 1.000 \\
QV & & 0.320 & & 0.476 & & 0.581 & & 0.946 \\
QW & & 0.099 & & 0.228 & & 0.345 & & 0.817 \\
VV & & 0.102 & & 0.227 & & 0.338 & & 0.895 \\
VW & & 0.032 & & 0.109 & & 0.200 & & 0.773 \\
WW & & 0.010 & & 0.052 & & 0.119 & & 0.667 \\
\hline
TT & & 0.137 & & 0.249 & & 0.350 & & 0.838 \\
\hline\hline
\end{tabular}
\end{table}

\subsubsection{Thermal Sunyaev-Zel'dovich effect}

In the case of the tSZ effect, the frequency dependence is exactly known; 
see Table~\ref{tab:fa}.  Unfortunately this frequency dependence is 
extremely weak in the \WMAP bands, with $F_a(\nu_1)F_a(\nu_2)$ varying by 
only a factor of 0.667 from the QQ to WW bands.  Therefore we must resort 
to the angular dependence to separate tSZ from lensing.

In order to perform an analysis similar to that of Sec.~\ref{sec:psfg}, we
need to determine or place a bound on the galaxy-SZ power
cross-spectrum, and determine the maximum flux $|F|$ of a tSZ source (we
use $|F|$ since tSZ sources have negative net flux in the \WMAP bands).  
In principle, tSZ haloes can be extended, however since our lensing
estimator uses information from multipoles $l\sim 350$ (physical
wavenumber $k/a=0.7$~Mpc$^{-1}$ at $D_{AP}\ge 0.5$~Gpc), the haloes will
not be resolved. This argument of course only applies to the single-halo
contribution to Eq.~(\ref{eq:dq}).  The flux from a tSZ source is
\begin{eqnarray}
F &=& -f_bT_{CMB}{\sigma_T\over D_{AP}^2}\,
   {k_B\langle T_e\rangle\over m_ec^2}
   \, {M\over\mu_em_p} \left(4-x\coth{x\over 2}\right)
\nonumber \\
&=& -8\times 10^{-4}\mu{\rm K~sr}\, \left({M\langle T_e\rangle\over 
   10^{16}M_\odot\,\rm keV}\right)
   \left( {f_b/\mu_e\over 0.18}\right)
\nonumber \\ && \times
   \left( {D_{AP}\over\rm Gpc}\right)^{-2}
   \left(4-x\coth{x\over 2}\right),
\label{eq:flux-tsz}
\end{eqnarray}
where $\sigma_T$ is the Thomson cross secton, $k_B$ is Boltzmann's 
constant, $f_b=\Omega_b/\Omega_m$, $m_e$ and $m_p$ are the electron and 
proton masses, $\langle T_e\rangle$ is the mean electron temperature, 
$\mu_em_p$ is the baryonic mass per free electron, $D_{AP}$ is the 
physical angular diameter distance, $M$ is the total mass of the halo, and 
$x=h\nu/k_BT_{CMB}=\nu/57$~GHz.  The frequency-dependent factor 
$4-x\coth(x/2)$ ranges from $1.91$ (Q band) to $1.57$ (W band).  For tSZ 
sources that are physically associated with the LRGs (distance $z\ge 0.2$ 
or $D_{AP}\ge 0.5$~Gpc) we will have $|F|<6\times 10^{-3}\mu$K~sr even for 
extremely massive ($M\langle T_e\rangle=10^{16}M_\odot\,$keV) haloes.  If 
we estimate contamination to the galaxy-convergence correlation 
in analogy to Eq.~(\ref{eq:psco}), we find
\beq
\Delta C^{g\kappa}_l({\rm tSZ,~1~halo}) =
  {r_{PS}(l)\over R_l}F\Delta C^{gT}_l;
\eeq
if we take $|F|=6\times 10^{-3}\mu$K~sr, the limits on the contamination
from the Q-band correlation $C^{gT}_l$ are similar to the limits for point
sources from $C^{gT}_l(Q)$ (cf. Fig.~\ref{fig:pscl}a).  Unfortunately, 
like our similar analysis for point sources, this analysis of the tSZ 
contamination does not tell us anything new, since if the contamination 
from tSZ in our $b_g$ measurement were large compared to the statistical 
error of $\pm 1.92$, we would have measured the wrong $b_g$.  Worse, it 
only applies to the single-halo contribution in Eq.~(\ref{eq:dq}), whereas 
the two-halo contribution is likely to dominate on sufficiently large 
scales.

\subsection{Galactic foregrounds}

Foreground microwave emission from our own Galaxy can introduce spurious
features in the weak lensing map.  Because of their Galactic origin,
these features cannot be correlated with the LRG distribution.  However,
it is possible that they can correlate with systematic errors in the LRG
maps, most notably (i) stellar contamination of the LRG catalog and (ii)
incomplete correction (or over-correction) for dust extinction.  We have
used two methods to address these potential problems.  The first
(Sec.~\ref{sec:gf-lrg}) is to correlate the CMB lensing map ${\bf v}$ with
stellar density and reddening maps.  The second (Sec.~\ref{sec:gf-cmb}) is
to correlate the LRG density map with simulated lensing contamination maps
$\Delta{\bf v}$ obtained by feeding microwave foreground templates through
the lensing pipeline.

\subsubsection{Stellar density and reddening tests}
\label{sec:gf-lrg}

The dominant systematics in the SDSS that could correlate with Galactic
microwave foregrounds are stellar contamination of the LRG catalog and
dust extinction.  We study these by constructing two maps: a map from SDSS
of the density $\delta n_\star/\bar n_\star$ of ``stars'' (defined as
objects with magnitude $18.0<r<19.5$ that are identified as pointlike by
the SDSS photometric pipeline \cite{2001adass..10..269L}), and a dust
reddening map of $E(B-V)$ from Ref.~\cite{1998ApJ...500..525S}.  These
maps can be substituted in place of the LRG map $\delta n_g/\bar n_g$ as
${\bf x}_{LRG}$ in Eq.~(\ref{eq:qa}) and the cross-spectrum and ``bias''
determined.  The biases obtained using these contaminant maps are shown in
Table~\ref{tab:contam}.

\begin{table}
\caption{\label{tab:contam}The bias obtained by substituting stellar 
density ($\delta n_\star/\bar n_\star$) and extinction [$E(B-V)$] maps in 
place of the LRG map ($\delta n_g/\bar n_g$), for the Kp05\istenpst sky 
cut.  The final column, $\Delta b_g$, is the error in the galaxy bias if 
we use the coefficients $c_\star$ and $c_E$ from the linear fit 
(Eq.~\ref{eq:linfit}) to estimate the contamination of the LRG map by 
stars and extinction.} 
\begin{tabular}{ccrcrcr} \hline\hline
Frequency & & stars & & $E(B-V)$ & & $\Delta b_g$ \\
\hline
QQ & & $-14.38$ & & $-0.17$ & & $+0.02$ \\
QV & &  $+2.37$ & & $+0.19$ & & $+0.10$ \\
QW & &  $-2.12$ & & $+0.01$ & & $+0.03$ \\
VV & &  $+3.40$ & & $+0.13$ & & $+0.05$ \\
VW & &  $-2.29$ & & $+0.14$ & & $+0.11$ \\
WW & &  $+3.99$ & & $+0.24$ & & $+0.11$ \\
\hline
TT & &  $+0.71$ & & $+0.15$ & & $+0.09$ \\
\hline\hline
\end{tabular}
\end{table}

A crude estimate of how this contamination translates into contamination 
of the galaxy-convergence power spectrum $C^{g\kappa}_l$ is provided by 
performing an unweighted least-squares fit of the LRG density to the 
reddening and stellar maps,
\begin{equation}
{\delta n_g\over\bar n_g} = c_EE(B-V) + c_\star{\delta n_\star\over\bar 
n_\star} + c_1 + {\rm residual},
\label{eq:linfit}
\end{equation}
over the 296,872 SDSS pixels.
The fit coefficients are $c_E=0.62$, $c_\star = -0.0092$, and 
$c_1=-0.017$.  We have shown in Table~\ref{tab:contam} the spurious 
contribution $\Delta{b_g}$ to the bias resulting from stellar and 
reddening contamination if one assumes these fit coefficients.

\subsubsection{Microwave foreground template test}
\label{sec:gf-cmb}

Since the lensing map ${\bf v}$ is a quadratic function of 
temperature, and the Galactic foregrounds are not correlated with the 
primary CMB, Eq.~(\ref{eq:dq}) is applicable to Galactic foregrounds.  The 
contamination $\Delta\langle{\bf v}^{\alpha\beta}\rangle$ can be obtained 
straightforwardly since the right-hand side of Eq.~(\ref{eq:dq}) can be 
evaluated by substituting in the foreground maps as $\Delta T^\alpha$.  
The difficult step is to construct a good foreground map $\Delta 
T^\alpha$; here we use external templates to avoid any possiblity of 
spurious correlations of the templates with the \WMAP data (either CMB 
signal or noise).

The Galactic foregrounds that must be considered in producing a template
at higher (W band) frequencies are free-free and thermal dust emission; at
lower frequencies (Q and V bands) an additional component is present whose
physical origin remains uncertain but which may include hard synchrotron
emission \cite{2003ApJS..148...97B} or spinning or magnetic dust
\cite{2003astro.ph.11547F, 2004ApJ...606L..89D}.  We have used Model 8 of
Refs. \cite{1998ApJ...500..525S, 1999ApJ...524..867F} for thermal dust,
and the H$\alpha$ line radiation template of Ref.
\cite{2003ApJS..146..407F} re-scaled using the conversions of Ref.
\cite{2003ApJS..148...97B} for free-free radiation.  There are no all-sky
synchrotron templates at the frequencies and angular scales of interest
(the Haslam radio continuum maps at 408 MHz \cite{1981A&A...100..209H,
1982A&AS...47....1H}, frequently used as a synchrotron template for CMB
foreground analyses, have a 50 arcmin FWHM beam and hence do not resolve
the $l\sim 200$--$400$ scales used for lensing of the CMB).  
Nevertheless, inclusion of the low-frequency component (whatever its
origin) is not optional, and so we follow Ref. \cite{2003astro.ph.11547F}
in modeling it as proportional to the thermal dust prediction of Ref.
\cite{1999ApJ...524..867F} multiplied by $T_{dust}^2$ using the
coefficients of Ref. \cite{2003astro.ph.11547F}.

As a test for contamination, we have substituted these foreground
templates for the true CMB maps, run them through the lensing pipeline,
and derived $b_g$ estimates by correlating against the true LRG map; the
results are shown in the ``foreground'' column of Table~\ref{tab:bias}.  
The typical contamination due to foregrounds is clearly very small (bias
error of a few times $10^{-4}$), and thus is negligible even if the
foreground amplitude has been underestimated by an order of magnitude (the
error on $b_g$ scales as the foreground amplitude squared).  This is not
surprising: in the relatively clean regions of sky used for this analysis,
the galactic foreground temperature anisotropy is roughly 2 orders of
magnitude (in amplitude) below the CMB temperature at the degree angular
scales.  Therefore a quadratic statistic, such as the lensing estimator,
should be $\sim 4$ orders of magnitude smaller than the foregrounds
(again, in amplitude).  Thus when ${\bf v}$(foreground) is correlated
against the galaxy map, the correlation that one expects from chance
alignments of foregrounds and galaxies is roughly 4 orders of magnitude
less than ${\bf v}$(CMB) (although a much greater correlation could 
exist if the galaxy map were also contaminated by foregrounds, e.g. dust 
extinction).  This is in contrast to the point sources, which are highly 
localized objects that become more and more dominant when we consider 
higher-order statistics such as the lensing estimator ${\bf v}$.

\section{Discussion}
\label{sec:discussion}

In this paper, we have carried out an initial search for weak lensing of
the CMB by performing a lensing reconstruction from the \WMAP data and
correlating the resulting lensing field map with the SDSS LRG map.  We do
not have a detection, however our result $b_g=1.81\pm 1.92$ ($1\sigma$) is
consistent with the bias $b_g\sim 1.8$ obtained from the LRG
clustering autopower \cite{padauto}.

The main purpose of this analysis was to identify any systematics that
contaminate the galaxy-convergence correlation at the level of the current
CMB data.  The good news is that our result for the bias is reasonable,
suggesting that such systematics are at most of the order of the
statistical errors.  We have also found that the Galactic foregrounds are
a negligible contaminant to the lensing signal (again, at the level of the
present data).  The bad news primarily concerns extragalactic foregrounds:
a significant amount of solid angle -- $21\%$ of Kp05\isten -- was lost
due to point source cuts that are necessary to avoid spurious power (at
least in Q band), and we have no assurance that significant point source
or tSZ contamination of the lensing signal does not lie just below the
threshold of detectability.  The extragalactic foreground analyses of
Sec.~\ref{sec:psfg} based on the frequency dependence of the signal and
the galaxy-temperature correlations yielded only a weak constraint on the
synchrotron point source contamination, $\Delta b_g^{(PS)}(TT) = 0.73\pm
1.18$, and essentially no useful constraint can be derived for tSZ using
the first-year \WMAP data.  Our constraints based on $C^{gT}_l$ for the
point sources are more stringent, $\Delta b_g^{(PS)} = -0.14\pm 0.51$, but 
these
assume Poissonianity of the sources, which must break down at some level.
The point source and tSZ issues will become even more important as future
experiments probe lensing of the CMB using higher-$l$ primary modes, where
point source and tSZ anisotropies contribute a greater fraction of the
total power in the CMB, and precision cosmology with lensing of the CMB
will require a means of constraining these contaminants in order to
produce reliable results.  Because in the real universe the extragalactic
foregrounds will not be exactly Poisson-distributed, the frequency
(in)dependence of the lensing signal will in principle provide the most
robust constraints on the contamination.  In this paper, we were unable to
obtain useful constraints this way because of the limited range of
frequencies on \WMAP (all in the Rayleigh-Jeans regime where the tSZ
signal has the same frequency dependence as CMB) and the low statistical
signal-to-noise.  Both of these problems should be alleviated with
high-resolution data sets covering many frequencies, e.g. as expected from
the {\slshape Planck} satellite \footnote{URL: {\tt
http://astro.estec.esa.nl/Planck}}.

The large solid angle that was lost to point source cuts in the analysis
presented here resulted from the need to remove ``artifacts'' in the ${\bf
v}$ map that occur around point sources.  One approach to this problem
would be to try to devise a CMB lensing reconstruction technique that
works with complicated masks.  Alternatively, one could compute the galaxy
density--CMB temperature--CMB temperature bispectrum
$B^{gTT}_{l_1l_2l_3}$, rather than trying to use the lensing field (a
quadratic function of CMB temperature) as an intermediate step; this way
one could mask out only the point source itself and not a 2 degree
exclusion radius around it.  The bispectrum approach carries the
additional advantage of retaining angular information about the
foregrounds; this information may be useful for separating lensing from
the kinetic SZ and patchy-reionization anisotropies that have no frequency
depedence but can still contaminate lensing if small-scale information is
used \cite{2003ApJ...598..756S, 2004astro.ph..2004V, 2004astro.ph..3075A}.  
The bispectum $B^{gTT}_{l_1l_2l_3}$ may therefore be of particular
interest for lensing analyses of high-$l$ experiments such as the Atacama
Pathfinder Experiment \footnote{URL: {\tt
http://bolo.berkeley.edu/apexsz/}}, the Atacama Cosmology Telescope
\footnote{URL:\\ {\tt
http://www.hep.upenn.edu/$\sim$angelica/act/act.html}}, and the South Pole
Telescope \footnote{URL: {\tt http://astro.uchicago.edu/spt/}}.

In summary, this paper represents a first analysis of lensing of the CMB
using real data, and should not be regarded as the last word on the
methodology.  We have identified extragalactic foregrounds (point sources
and tSZ) as the most worrying contaminant to the lensing signal in the
\WMAP data; the point sources, if unmasked, dominate the power spectrum of
the reconstructed convergence if the Q band data are used, but this effect
is suppressed at higher frequencies.  We have shown that in the current
data, the Galactic foreground contribution is negligible, and the
contamination from point sources and tSZ in the galaxy-convergence
cross-spectrum is at most of order the signal (although we have no
detection of contamination).  Like the search for the CMB lensing signal
(and its eventual use in precision cosmology), stronger statements about
the foreground contamination must await higher signal-to-noise data at a
wide range of frequencies.

\acknowledgments

We acknowledge useful discussions with Niayesh Afshordi, Joseph Hennawi,
Yeong-Shang Loh, and Lyman Page. C.H. is supported through NASA grant
NGT5-50383.  U.S. is supported by Packard Foundation, NASA NAG5-11489, and
NSF CAREER-0132953.

Some of the results in this paper have been derived using the {\sc
HEALPix} \cite{1999elss.conf...37G} package. We acknowledge the use of the
Legacy Archive for Microwave Background Data Analysis (LAMBDA)  
\footnote{URL: {\tt http://lambda.gsfc.nasa.gov/}}. Support for LAMBDA is
provided by the NASA Office of Space Science.  We used computational
resources provided by NSF grant AST-0216105.

Funding for the creation and distribution of the SDSS Archive has been
provided by the Alfred P. Sloan Foundation, the Participating
Institutions, the National Aeronautics and Space Administration, the
National Science Foundation, the U.S. Department of Energy, the Japanese
Monbukagakusho, and the Max Planck Society. The SDSS Web site is {\tt
http://www.sdss.org/}.

The SDSS is managed by the Astrophysical Research Consortium (ARC) for the
Participating Institutions. The Participating Institutions are The
University of Chicago, Fermilab, the Institute for Advanced Study, the
Japan Participation Group, The Johns Hopkins University, Los Alamos
National Laboratory, the Max-Planck-Institute for Astronomy (MPIA), the
Max-Planck-Institute for Astrophysics (MPA), New Mexico State University,
University of Pittsburgh, Princeton University, the United States Naval
Observatory, and the University of Washington.

\appendix

\section{Non-isolatitude spherical harmonic transform}
\label{app:nilsht}

The non-isolatitude spherical harmonic transform (SHT) is used in our
cross-correlation analysis.  The SHT operations on the unit sphere 
transform
between real- and harmonic-space representations of a function:
\beq
\begin{array}{lcl}
T(\nhat_i) = \sum_{l=0}^{l_{\rm max}} \sum_{m=-l}^l T_{lm} 
Y_{lm}(\nhat_i),
& &
{\rm (synthesis)}
\\
S_{lm} = \sum_{i=0}^{N_p-1} Y_{lm}^\ast(\nhat_i) S(\nhat_i).
& &
{\rm (analysis)} \end{array}
\label{eq:sht-s}
\eeq
For high-resolution data sets, this operation is usually performed using
an isolatitude pixelization, i.e. one in which the pixels are positioned
on curves of constant colatitude $\theta$.  This situation allows the
colatitude ($\theta$) and longitude ($\phi$) parts of the spherical
transform to be performed independently, resulting in an overall operation
count scaling as $O(N_p^{3/2})$
\cite{1999elss.conf...37G,1998astro.ph..6374C}.  While this approach
works, and has contributed remarkably to the popularity of isolatitude
pixelizations such as HEALPix, there are reasons to maintain the
flexibility to use any pixels.  For example, in simulations of
gravitational lensing of the CMB, we need to produce a simulated lensed
map, and in
general a set of pixels that are isolatitude in ``observed'' coordinates
(e.g. HEALPix) maps onto a non-isolatitude grid on the primary CMB.  We
note that for other analyses there may be other reasons to consider more
general pixelizations, which preserve desired properties such as
conformality \cite{2003astro.ph..3020C} or maximal symmetry
\cite{1996ApJ...470L..81T}. This Appendix describes our non-isolatitude
SHT algorithm.

\subsection{The method}

We consider first the SHT synthesis.  Our first step is to perform a
latitude transform using associated Legendre
polynomials on a set of points equally spaced in $\theta$ (the ``coarse
grid''):
\begin{equation}
T_m({\alpha\over L}\pi) = \sum_{l=|m|}^\lmax T_{lm} Y_{lm}(\theta =
{\alpha\over L}\pi,\phi=0).
\label{eq:coarse}
\end{equation}
This procedure is performed for integers in the range $0\le\alpha\le L$,
$-\lmax\le m\le\lmax$.  [Here $L$ is an integer satisfying $L>\lmax$,
which we require to be a power of two times a small odd integer.  The
first requirement ensures that Eq. (\ref{eq:coarse}) over-Nyquist samples
the variations in $T_m(\theta)$, the second ensures that the FFT is a fast
operation.] It requires a total of $O(\lmaxtwo L)$ operations and is the
most computationally demanding step in the transform.  The spherical
harmonics are computed as needed using an ascending recursion relation.
The standard recursion relation is used to generate the associated
Legendre functions; we speed up the transform by a factor of two over the
brute-force approach by taking advantage of the symmetry/antisymmetry of
the spherical harmonics across the equatorial plane.

The next step is to refine the coarse grid, which has a spacing of
$\pi/L$ in $\theta$, to a ``fine grid'' with spacing
$\pi/L'$ where $L'>L$.  We do this
by taking advantage of the band-limited nature of the spherical
harmonics.  Any linear combination of spherical
harmonics of order $l\le\lmax$ can be written as a band-limited function:
\begin{equation}
T_m(\theta) =
\sum_{l=|m|}^\lmax T_{lm} Y_{lm}(\theta,0) = \sum_{n=-\lmax}^\lmax
C_{m,n} e^{in\theta}.
\label{eq:bandlimit}
\end{equation}
We may determine the coefficients $C_{m,n}$ via a fast Fourier transform
(FFT) of length $2L$, so long as the left-hand side has been evaluated at
the points $\theta = \pi\alpha/L$ for integers $-L<\alpha\le L$.  (We use
parity rules to compute the left hand side at negative values of
$\theta$.)  By applying an FFT of length $2L'$ to the $C_{m,n}$, we then
obtain $T_m(\theta)$ at values of $\theta = \pi\alpha/L'$ for integers
$0\le\alpha\le L'$.  What has been gained here is that we have performed
the associated Legendre transform on $2L'$ points, but the expensive
evaluation of the associated Legendre polynomials has only been required
at $2L$ points.  If $L$ and $L'$ are powers of two, then the FFT process
requires $O(\lmax L'\log L')$ operations.

The third step is an FFT in the longitude direction to obtain
$T(\theta=\pi\alpha/L',\phi=\pi\gamma/L')$, where $\alpha$ and $\gamma$
are integers.  This process has become standard in isolatitude SHT
algorithms; in its full glory, it is given by:
\begin{equation}
T[\alpha,\gamma]\equiv
T(\theta={\alpha\over L'}\pi,\phi={\gamma\over L'}\pi)
= \sum_{m=-\lmax}^\lmax T_m(\theta) e^{im\phi}.
\end{equation}
At the end of this step, we know the real-space value of our function on a
fine equicylindrical projection (ECP) grid of spacing $\pi/L'$.  The FFT
operation in this step requires $O(L'^2\log L')$ operations.  The total
operation count of transforming onto the ECP grid is $O(\lmaxthree)$ and
is dominated by the associated Legendre transform, so long as $L'/\lmax <
\sqrt{\lmax}/\log\lmax$.  What is important to note is that, particularly
if $\lmax$ is large (for a future CMB polarization experiment we have to
consider multipoles up to roughly $\lmax\sim 4000$), we can sample our
function in real space at several times the Nyquist frequency at no
additional computational cost.  This is exactly what is required in order
to successfully interpolate the values $T_j$.

The final step is the interpolation step.  For each point $\nhat_j$, we
identify the coordinates in the ECP grid: (we suppress the $j$ index here
for clarity)
\begin{equation}
\alpha + \delta_\alpha = L'{\theta\over\pi};\;
\gamma + \delta_\gamma = L'{\phi\over\pi},
\end{equation}
where $\alpha$ and $\gamma$ are integers and the fractional parts
$0\le\delta_\alpha,\delta_\gamma<1$. A $4K^2$-point, two-dimensional
polynomial interpolation is then computed:
\begin{equation}
T\approx \sum_{\mu = -K+1}^K w_\mu(\delta_\alpha) \sum_{\nu = -K+1}^K
w_\nu(\delta_\gamma) T[\alpha+\mu,\gamma+\nu],
\label{eq:interp}
\end{equation}
where the weights $w_\rho(\delta)$ are computed by Lagrange's formula:
\begin{equation}
w_\rho(\delta) = \frac{(-1)^{K-\rho}}{(K-\rho)!(K-1+\rho)!(\delta-\rho)}
\prod_{\sigma=-K+1}^K (\delta-\sigma).
\label{eq:weights}
\end{equation}
The weights for both the $\alpha$ and $\gamma$ directions may be evaluated
in a total of $O(K)$ multiplications and divisions if the factorials have
been pre-computed, so that for high-order interpolations the dominant
contribution to the computation time in interpolation comes from the
multiplications in Eq. (\ref{eq:interp}) rather than from computation of
the weights.

Note that the ``analysis'' operation of Eq. (\ref{eq:sht-s}) is the matrix
transpose operation of the ``synthesis'' (if we view the pixelized
$T(\nhat_i)$ and the harmonic-space $T_{lm}$ as vectors), and that all of
the steps outlined above are linear operations.  Since the transpose of a
composition of operations is the composition of the transposes in reverse
order, we can simply use the transposes of these operations in reverse
order to compute an SHT analysis using the same number of operations.

The above SHT algorithm generalizes easily to vector spherical 
harmonics, for which we use the basis:
\beqa
Y_{lm}^{(\parallel)}(\nhat) = && \!\!\!\!
{1\over\sqrt{l(l+1)}} \nabla Y_{lm}(\nhat) , \nonumber \\
Y_{lm}^{(\perp)}(\nhat) = && \!\!\!\!
{1\over\sqrt{l(l+1)}} \nhat\times\nabla Y_{lm}(\nhat).
\label{eq:sht-v}
\eeqa

\subsection{Interpolation accuracy}

The order $K$ and step size $\pi/L'$ of the interpolating polynomial is
determined by a balance of computation time, memory usage, and
accuracy.  The computational cost of the interpolation is $O(K^2N_{pix})$
and the memory usage is $O({L'}^2)$.  Hence it is important to understand
how interpolation accuracy is related to $K$ and $L'$.

\begin{figure}
\includegraphics[angle=-90,width=3in]{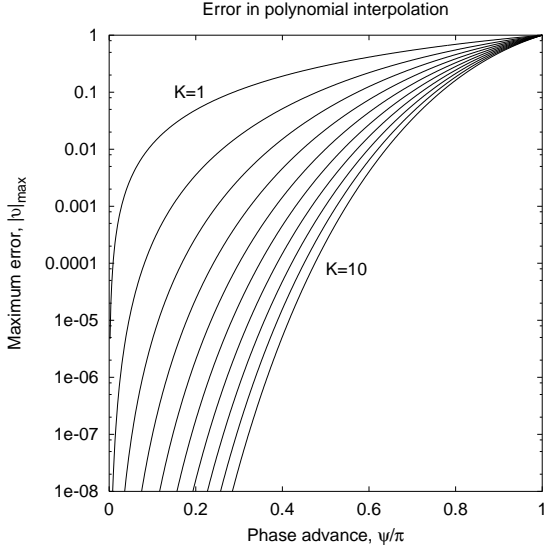}
\caption{\label{fig:upsilon}The maximum fractional error $|\upsilon_{\rm
max}|$ from Eq.~(\ref{eq:upsilon-bound}) as a function of the
interpolation order $K$ and the sampling rate.  From top to bottom, the
curves correspond to $K=1$ to $K=10$.  Note that $\psi=\pi$ for data
sampled at the Nyquist frequency (i.e. $L'=\lmax$), whereas $\psi=\pi/2$
for data sampled at twice the Nyquist frequency ($L'=2\lmax$),
etc.  Accuracy can be improved via either fine sampling (small $\psi$) or
high-order interpolation (high $K$).}
\end{figure}

The error in polynomial interpolation of $T(\theta,\phi)$ can be
determined from the error in interpolation of a band-limited
function.  We note that the error in interpolation of a given Fourier mode
$f(\theta,\phi)=C_{m,n}e^{i(n\theta+m\phi)}$ is given by:
\beq
\frac{f_{interp}(\theta,\phi)}{f(\theta,\phi)} =
\left[ 1 + \upsilon\left(\delta_\alpha, 2\pi{m\over L'}\right) \right]
\left[ 1 + \upsilon\left(\delta_\gamma, 2\pi{n\over L'}\right) \right],
\eeq
where the $\upsilon$ function is:
\beq
\upsilon(\delta, \psi) =
\sum_{\rho = -K+1}^K w_\rho(\delta) e^{i\psi(\rho-\delta)} - 1.
\eeq
(Note that $\psi$ represents the phase advance of our Fourier mode per
grid cell.)  The $\upsilon$ function determines the fractional error in a
given Fourier mode.  It is most easily evaluated by noting that (since
polynomial interpolation is exact for a constant) we have
$\upsilon(\delta,0)=0$, and derivative:
\beq
\left|{\partial\upsilon\over\partial \psi}\right| =
\frac{ \prod_{\sigma=-K+1}^K |\delta-\sigma| }
{(2K-1)!} |1 - e^{-i\psi}|^{2K-1}.
\eeq
[Here we have used Eq.~(\ref{eq:weights}) and applied the
binomial theorem.]  We note that the product is maximized at $\delta=1/2$
and that using trigonometric identities the exponential term can be
simplified.  Since $\upsilon(\delta,0)=0$, we have that $|\upsilon|$
cannot exceed the integral of $|d\upsilon/d\psi|$ from $0$ to $\psi$:
\beq
|\upsilon(\delta,\psi)| \le {(2K)!\over
4^KK!(K-1)!} \int_0^\psi \sin^{2K-1} {\psi'\over 2} d\psi'.
\label{eq:upsilon-bound}
\eeq
This bound is plotted in Fig.~\ref{fig:upsilon}.

\section{Preconditioner}
\label{app:prec}

One of the steps in the cross-spectrum estimator requires the solution of 
the linear system:
\beq
{\sf C}_p{\bf y} = {\bf x}
\label{eq:cy}
\eeq
for ${\bf y}$.  Here ${\sf C}_p$ is the prior covariance matrix for the 
LRGs and is
given by ${\sf C}_p={\sf S}_p+{\sf N}$, where ${\sf S}_p$ is the signal
prior and ${\sf N}$ is the noise.  Since the matrix ${\sf C}_p$ is too
large ($N(N+1)/2$ elements where the dimension $N=296,872$) to
store in memory, Eq.~(\ref{eq:cy}) must be solved using iterative methods.
We have used a preconditioned conjugate-gradient (PCG) method.
The PCG method \cite{1992nrca.book.....P, 1999ApJ...510..551O} with 
preconditioner ${\sf E}$ produces successive estimates ${\bf y}^{(i)}$ for 
the solution to Eq.~(\ref{eq:cy}) using the equation:
\beq
{\bf y}^{(i)} = {\bf y}^{(i-1)} + {{\bf r}^{(i-1)T}{\sf E}{\bf r}^{(i-1)}
\over {\bf p}^{(i)}{\sf C}_p{\bf p}^{(i)}} {\bf p}^{(i)},
\eeq
where the residuals are defined by ${\bf r}^{(i)} = {\bf x}-{\sf 
C}_p{\bf y}^{(i)}$, and the search directions ${\bf p}^{(i)}$ are 
chosen according to
\beq
{\bf p}^{(i)} = {\sf E}{\bf r}^{(i-1)} +
{ {\bf r}^{(i-1)T}{\sf E} {\bf r}^{(i-1)} \over
  {\bf r}^{(i-2)T}{\sf E} {\bf r}^{(i-2)} } {\bf p}^{(i-1)}.
\eeq
The initial conditions are ${\bf y}^{(0)}=0$ and ${\bf p}^{(1)}={\sf
E}{\bf x}$.

The choice of preconditioner strongly affects the rate of convergence of
the PCG algorithm; ideally we have a preconditioner for which ${\sf E}{\bf
u}$ can be rapidly computed for given ${\bf u}$, and for which ${\sf
E}\approx{\sf C}_p^{-1}$.  In using PCG for CMB power spectrum estimation,
Ref.~\cite{1999ApJ...510..551O} used a preconditioner based on
(approximate) azimuthal symmetry of the Galactic plane cut.
Unfortunately, the SDSS survey region lacks any such symmetry.
Therefore we use a preconditioner that works independently of
any symmetries of the survey region.  The strategy is to use a two-scale
preconditioner: the low-resolution scale ($l<l_{split}$) is solved by
brute-force matrix inversion in harmonic space, while the high-resolution
scale ($l\ge l_{split}$) is essentially un-preconditioned and the burden
of convergence lies on the conjugate-gradient algorithm.  This is useful
because the large condition ratio of a typical prior matrix ${\sf C}_p$
comes almost entirely from a few large eigenvalues corresponding to the
low $l$ modes.  The condition ratio is dramatically improved by
suppressing these eigenvalues with the two-scale preconditioner.

The preconditioner is obtained by cutting the power spectrum at
$l_{split}$ and defining the $l_{split}^2\times l_{split}^2$ matrix ${\sf
M}$:
\beqa
M_{lm,l'm'} &=& \delta_{ll'}\delta_{mm'}
\nonumber \\ && +
\frac{\sqrt{(S_{p,l}-S_{p,l_{split}})(S_{p,l'}-S_{p,l_{split}})}}
{S_{p,l_{split}}}
\nonumber \\ && \times
\sum_j \Omega_j Y_{lm}^\ast(\nhat_j) Y_{l'm'}(\nhat_j).
\label{eq:tlmlm}
\eeqa
Here $l$ and $l'$ are in the range from $0$ to $l_{split}-1$, $j$ is a
pixel index, and $\Omega_j$ is the area of the $j$th pixel.  Note that a
red prior spectrum (i.e. $S_{p,l}>S_{p,l_{split}}$) has been assumed, as
appropriate for the angular power spectra of galaxies.  We next define the 
${\sf G}$ matrix:
\beqa
G_{lm,l'm'} &=&
\frac{\sqrt{(S_{p,l}-S_{p,l_{split}})(S_{p,l'}-S_{p,l_{split}})}}
{S_{p,l_{split}}}
\nonumber \\ && \times [{\sf M}^{-1}]_{lm,l'm'}.
\label{eq:l-pre}
\eeqa
Then the preconditioner is:
\beqa
E_{ij} &=& \frac{\Omega_i\delta_{ij}}{S_{p,l_{split}}}
- \sum_{lml'm'} \frac{\Omega_i\Omega_j}{S_{p,l_{split}}^2}
\nonumber \\ && \times
Y_{lm}(\nhat_i) Y_{l'm'}^\ast(\nhat_j) G_{lm,l'm'}.
\label{eq:e-pre}
\eeqa

Using the Woodbury matrix inverse formula, the matrix ${\sf E}$ of
Eq.~(\ref{eq:e-pre}) can be recognized as the exact inverse of the matrix:
\beqa
[{\sf E}^{-1}]_{ij} &=& S_{p,l_{split}}{\delta_{ij}\over \Omega_i}
+ \sum_{l=0}^{l_{split}-1}
(S_{p,l}-S_{p,l_{split}})
\nonumber \\ && \times
\sum_{m=-l}^l Y_{lm}(\nhat_i) Y_{lm}^\ast(\nhat_j).
\eeqa
That is, ${\sf E}$ is the exact inverse of a convolution that has
power spectrum $S_{p,l}$ for $l<l_{split}$ and white noise with power
$S_{p,l_{split}}$ for $l\ge l_{split}$.  Thus ${\sf E}$ is a good
approximation to ${\sf C}_p$ if $l_{split}$ is chosen to be
the multipole where the signal and noise have similar power.  In this
case, the convergence of the PCG iteration is very rapid.  In practice
(see below) we choose $l_{split}$ to be somewhat less than this in order
to speed up multiplication of a vector by ${\sf E}$, and accept that more
iterations will be required for convergence.

Note that ${\sf L}$ need only be computed once for each sky cut and prior
power spectrum; it can then be stored and used for many ${\sf E}{\bf u}$
operations.  The ${\sf E}{\bf u}$ operation is then reduced to a spherical
harmonic transform of cost $O(l_{split}^3)+O(K^2N_{pix})$, a matrix-vector
multiplication of computational cost $O(l_{split}^4)$, and another
spherical harmonic transform.  Since we use the preconditioner once for
every ${\sf C}_p$ operation, we gain speed by increasing $l_{split}$ until
the most expensive part of the ${\sf E}$ operation is of comparable cost
to the ${\sf C}_p$ operation ($O(\lmaxthree)$).  This suggests that we set
$l_{split}\sim l_{\rm max}^{3/4}$.  There are however practical
limitations on $l_{split}$: the size of the matrix in memory is
$O(l_{split}^4)$, and the computational cost of obtaining ${\sf G}$ is
$O(l_{split}^6)$.  Thus the best value of $l_{split}$ is generally
somewhat less than $l_{\rm max}^{3/4}$; we have used $l_{split}=32$.

\end{document}